\documentclass[twocolumn]{emulateapj}
\usepackage{color}
\usepackage{url}
\usepackage[usenames,dvipsnames,svgnames,table]{xcolor}
\usepackage[colorlinks=blue,
           linkcolor=blue,
           urlcolor=blue,
           citecolor=blue]{hyperref}

\newcommand{\kms}{km~s$^{-1}$}

\begin{document}
\title{Observation of a Short Period Quasi-Periodic Pulsation in Solar X-ray, Microwave and EUV Emissions}
\author{PANKAJ KUMAR\altaffilmark{1,2}, VALERY M. NAKARIAKOV\altaffilmark{3,4,5}, KYUNG-SUK CHO\altaffilmark{1,6}}
\affil{$^1$Korea Astronomy and Space Science Institute (KASI), Daejeon, 305-348, Republic of Korea}
\affil{$^2$Heliophysics Science Division, NASA Goddard Space Flight Center, Greenbelt MD 20771}
\affil{$^3$Centre for Fusion, Space and Astrophysics, Department of Physics, University of Warwick, CV4 7AL, UK}
\affil{$^4$School of Space Research, Kyung Hee University, Yongin, 446-701, Gyeonggi, Republic of Korea}
\affil{$^5$ St Petersburg Branch, Special Astrophysical Observatory, Russian Academy of Sciences, 196140, St Petersburg, Russia}
\affil{$^6$University of Science and Technology, Daejeon 305-348, Republic of Korea}
\email{pankaj@kasi.re.kr}

\begin{abstract}

This paper presents the multi-wavelength analysis of a 13~s quasi-periodic pulsation (QPP) observed in hard X-ray (12--300 keV) and microwave (4.9--34 GHz) emissions during a C-class flare occurred on 21 September 2015. 
AIA 304 and 171 \AA~ images show an emerging loop/flux tube (L1) moving radially outward, which interacts with preexisting structures within the active region. The QPP was observed during the expansion and rising motion of L1.   
The Nobeyama Radioheliograph (NoRH) microwave images in 17/34 GHz channels reveal a single radio source, which was co-spatial with a neighboring loop (L2). In addition, using AIA 304 \AA~ images, we detected intensity oscillations in the legs of loop L2 with a period of about 26~s. A similar oscillation period was observed in the GOES soft X-ray flux derivative. This oscillation period seems to increase with time. We suggest that the observed QPP is most likely generated by the interaction between loops L2 and L3 observed in the AIA hot channels (131 and 94 \AA). The merging speed of loops L2 and L3 was $\sim$35 \kms. Loop L1 destroyed possibly by its interaction with preexisting structures in the active region and produced a cool jet with the speed $\sim$106--118~\kms associated with a narrow CME ($\sim$770 \kms). Another mechanism of the QPP in terms of a sausage oscillation of the loop (L2) is also possible. 
\end{abstract}

\section{INTRODUCTION}
Quasi-periodic pulsations (QPPs) with typical periods ranging from a few seconds to several minutes are often observed in the light curves of solar and stellar flares, taken in optical, EUV, X-ray, and radio wavelengths \citep[e.g.][]{nakariakov2009,pandey2009, 2010SoPh..267..329K, 2015SoPh..290.3625S, pugh2015, inglis2015,cho2016}. The physical mechanism for the generation of QPPs is important for understanding the drivers of the flaring energy releases, particle acceleration, associated plasma heating, and may be useful for the flare forecasting. 

QPPs in solar flares are usually associated with several mechanisms: (i) periodic regime of spontaneous magnetic reconnection, associated with tearing of the current sheet  and/or the formation/ejection of multiple plasmoids \citep[\lq\lq magnetic dripping\rq\rq, see, e.g. ][]{kliem2000,karl2007,barta2007,barta2008,kumar2013p}, (ii) direct modulation of the flaring plasma or charged particle kinematics by magnetohydrodynamic (MHD) waves \citep[e.g.][]{zaitsev1982, 2006A&A...446.1151N}, (iii) magnetic reconnection induced periodically by MHD waves \citep{chen2006,nakariakov2006}, (iv) alternate current in an equivalent LCR-circuit \citep[e.g.][]{1998A&A...337..887Z}, and (v) oscillatory regime of the coalescence of current carrying loops \citep{tajima1987, 2016PhRvE..93e3205K}. Some of these mechanisms have been confirmed observationally.
For example, the leakage of umbral oscillations from sunspots as slow magnetoacoustic waves was found to produce 3--5 minute QPP in small-scale explosive events and flares \citep{ning2004,chen2006, sych2009}.
Recently, \citet{kumar2015,kumar2016} reported two interesting 3-min QPPs in X-ray, radio and EUV channels. These QPPs were found to correlate with the leakage of umbral waves from a nearby sunspot, and small untwisting filament interacting with the ambient field. Therefore, leakage of the sunspot umbral oscillation or an untwisting filament could trigger the repetitive reconnection at the magnetic null point in a fan-spine topology. 

QPPs observed in hard X-ray (HXR) associated with the non-thermal electrons with the energy greater than several tens of keV, and microwave, with the energies greater than hundreds of keV, are generally associated with the periodic acceleration of nonthermal electrons that then emit the observed radiation by bremsstrahlung and gyrosynchrotron mechanisms, respectively \citep[e.g.][]{dulk1985,asc1987}. The time profiles of HXR and microwave fluxes are usually seen to be well-correlated with each other, including QPP. It suggests that the same population of the energetic electrons is responsible for these signals \citep{nakajima1983,asai2001,grechnev2003,inglis2009}. In particular, the kink oscillation caused by a fast magnetoacoustic mode, of a coronal loop near the reconnection site, could periodically induce magnetic reconnection. Hence it could periodically modulate the associated periodic acceleration of charged particles,  driving HXR and microwave QPPs \citep[e.g.][]{foullon2005,nakariakov2006,inglis2009}. In particular, \citet{nakariakov2010q} reported a 40-s QPP detected simultaneously in phase at HXR, microwave and gamma-ray, and suggested triggering of repetitive reconnection by a kink oscillation of a nearby loop. Likewise, slow magnetoacoustic waves could periodically trigger magnetic reconnection, which, in particular, explains well the progression of the energy release site along the neutral line of the plasma arcade, typical for two-ribbon flares \citep{nakariakov2011}.

Sausage oscillations of thick/dense flare loops may cause the density perturbation and associated modulation of HXR and microwave emission during solar flares \citep{zaitsev1982,roberts1984}. Almost-decayless QPPs with a period of 0.5-60 s may be excited by the sausage oscillation of thick/fat loops. The theoretical condition for the existence of the trapped global sausage-mode oscillation is $n_o/n_e>2.4(l/w)^2$, where $n_o/n_e$ is the density contrast and $l/w$ is the ratio of the length and width of the loop.
Therefore, high-density contrast is required in the oscillating loop to satisfy the condition for the existence of a fundamental sausage mode with a high quality factor, which can be achieved in the case of flaring loops \citep{asc2004a}.  Using NoRH observations, \citet{nakariakov2003} detected 14--17~s QPPs, oscillating in phase at the loop apex and at its legs, and interpreted it in terms of the fundamental sausage mode. In loops with lower density contrasts sausage oscillations can exist too, but they are subject to leakage to the external medium, and hence have lower quality factors \citep{2012ApJ...761..134N}.

Previous studies of short period QPPs (in the SOHO/EIT era) generally lack the detailed study of the spatial structure of the QPP source, because of the unavailability of high-resolution imaging data. At present, the analysis of AIA (12-s cadence) extreme ultraviolet (EUV) images together with the hard X-ray (HXR) and microwave data provides an excellent opportunity to explore the relevant drivers of QPPs in more detail. 

In this paper, we report the detection of a 13-s QPP in a short duration C4.2 flare occurred in active region (AR) NOAA 12420 on 21 September 2015. The pulsation was observed mainly in the HXR and microwave channels. In addition, a longer period, 26-s QPP was detected in the EUV intensity and GOES soft X-ray flux derivative.  In section 2, we present the observations, and in the last section, we discuss and summarise the results.


\begin{figure*}[h]
\centering{
\includegraphics[width=7.3cm]{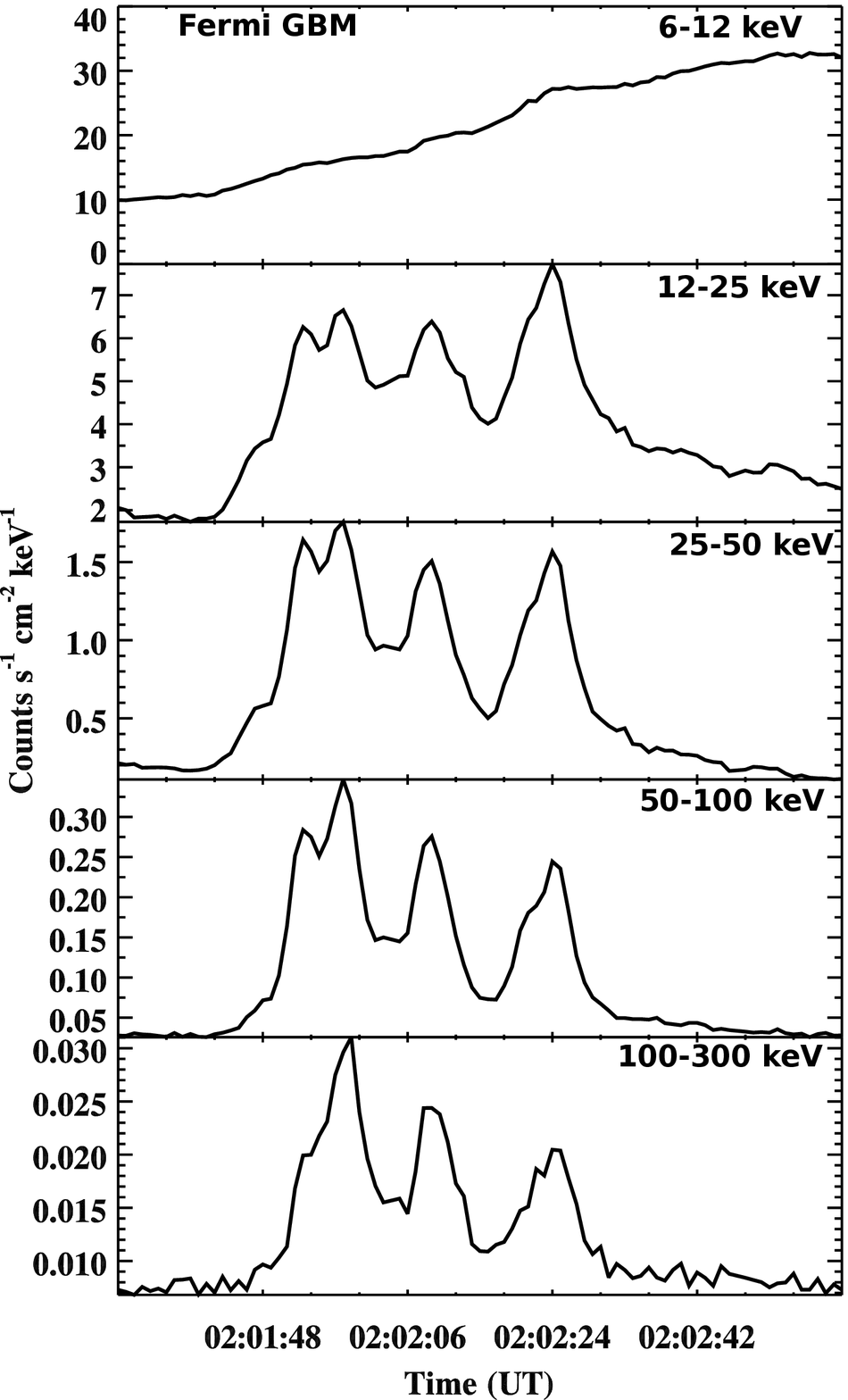}
\includegraphics[width=7.3cm]{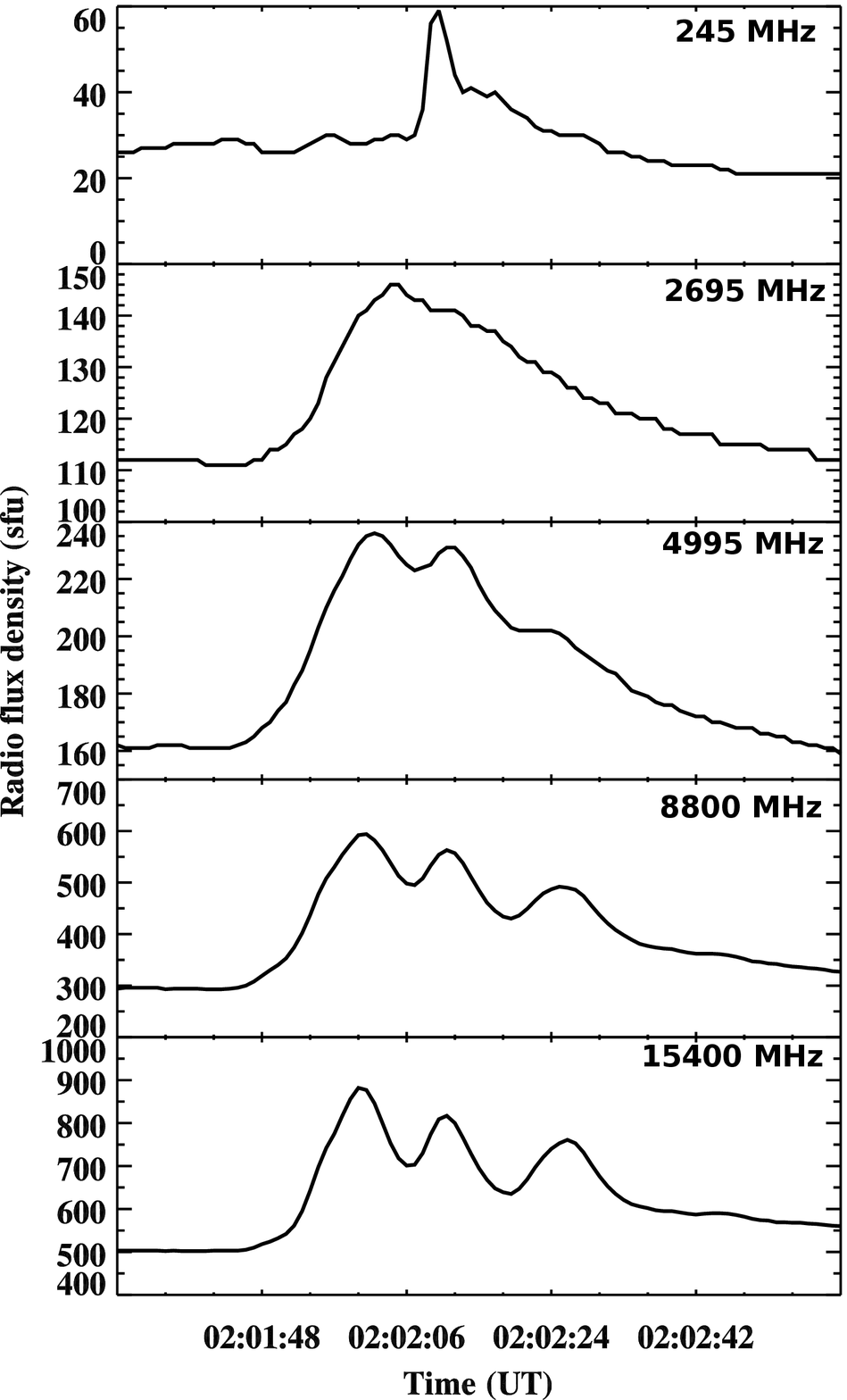}
}
\caption{Left: 4-s cadence X-ray flux profiles from Fermi GBM in 6-12, 12-25, 25-50, 50-100, and 100-300 keV channels. Right: 1-s cadence radio flux profiles (in sfu unit, 1 sfu=10$^{-22}$ W m$^{-2}$ Hz$^{-1}$) in 245, 2695, 4995, 8800, and 15400~MHz frequency bands observed at the RSTN Learmonth radio observatory. The short-period decaying oscillation is mainly observed in the microwave channels (4995, 8800, and 15400~MHz).} 
\label{fermi_rstn}
\end{figure*}

\begin{figure*}[h]
\centering{
\includegraphics[width=8.0cm]{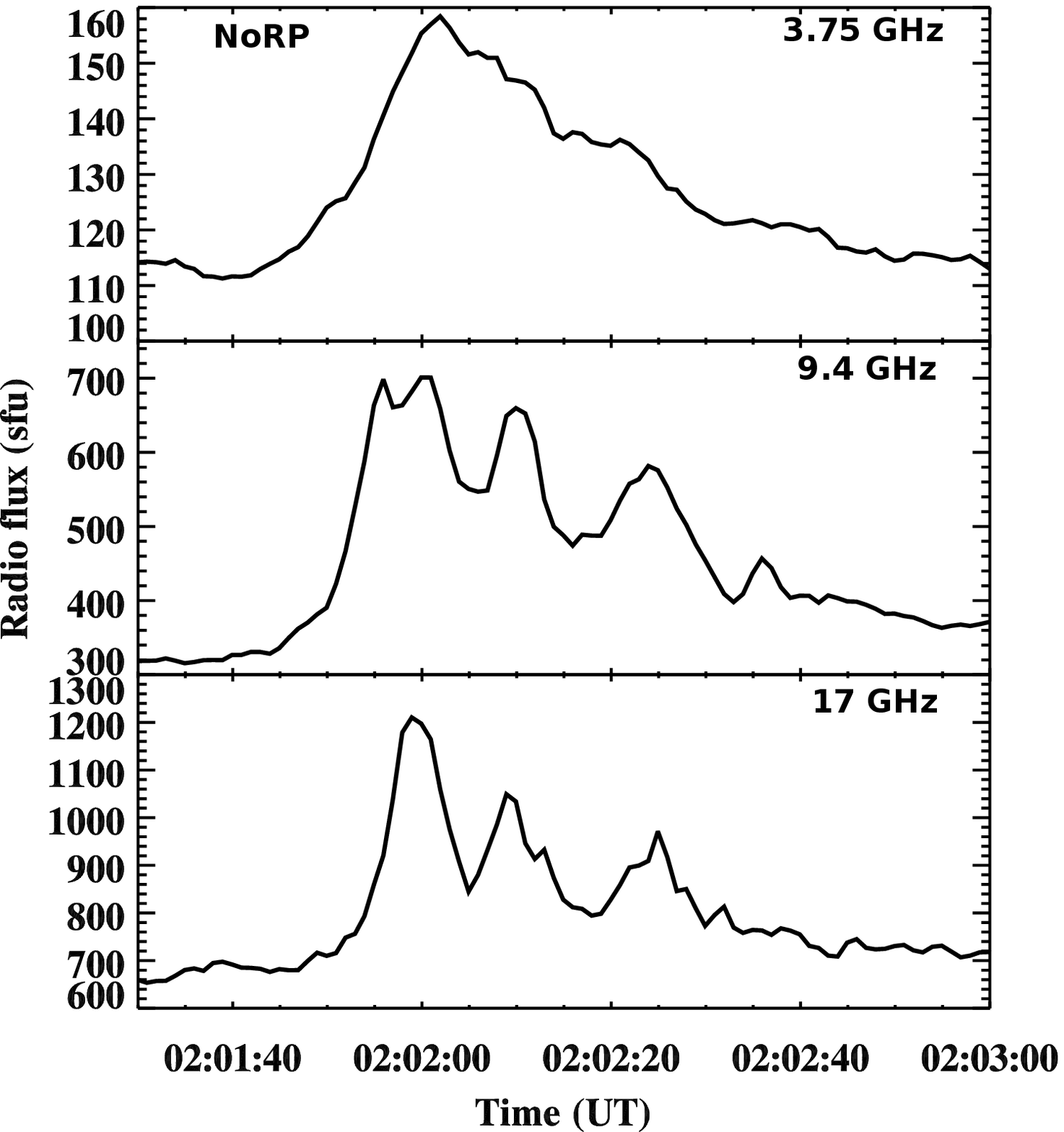}
\includegraphics[width=6.5cm]{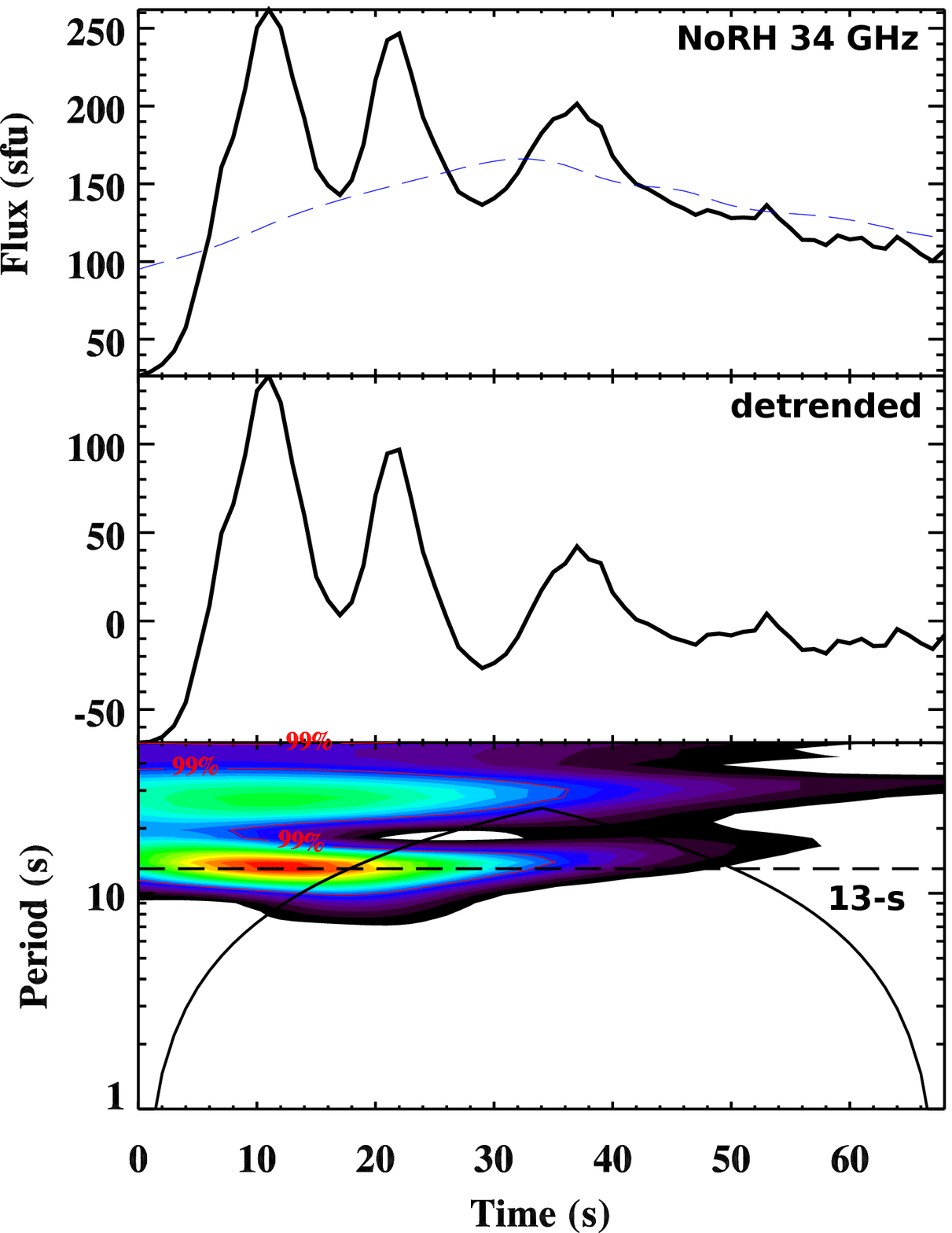}
}
\caption{{\it Left:} 1-s cadence radio flux profiles (in sfu unit, 1 sfu=10$^{-22}$ W m$^{-2}$ Hz$^{-1}$) in the 3.75, 9.4, and 17 GHz frequency bands observed by Nobeyama radio polarimeters (NoRP). The short-period decaying oscillation is observed in the hard X-ray and microwave channels. Four cycles with a decaying pattern are clearly observed in the 9.4~GHz channel. {\it Right:} 1-s cadence NoRH 34 GHz flux profiles. Wavelet power spectrum of the 34~GHz detrended signal. The period of oscillation is about 13 s. The start time is 02:01:48 UT.}
\label{nobeyama}
\end{figure*}

\begin{figure*}
\centering{
\includegraphics[width=6.5cm]{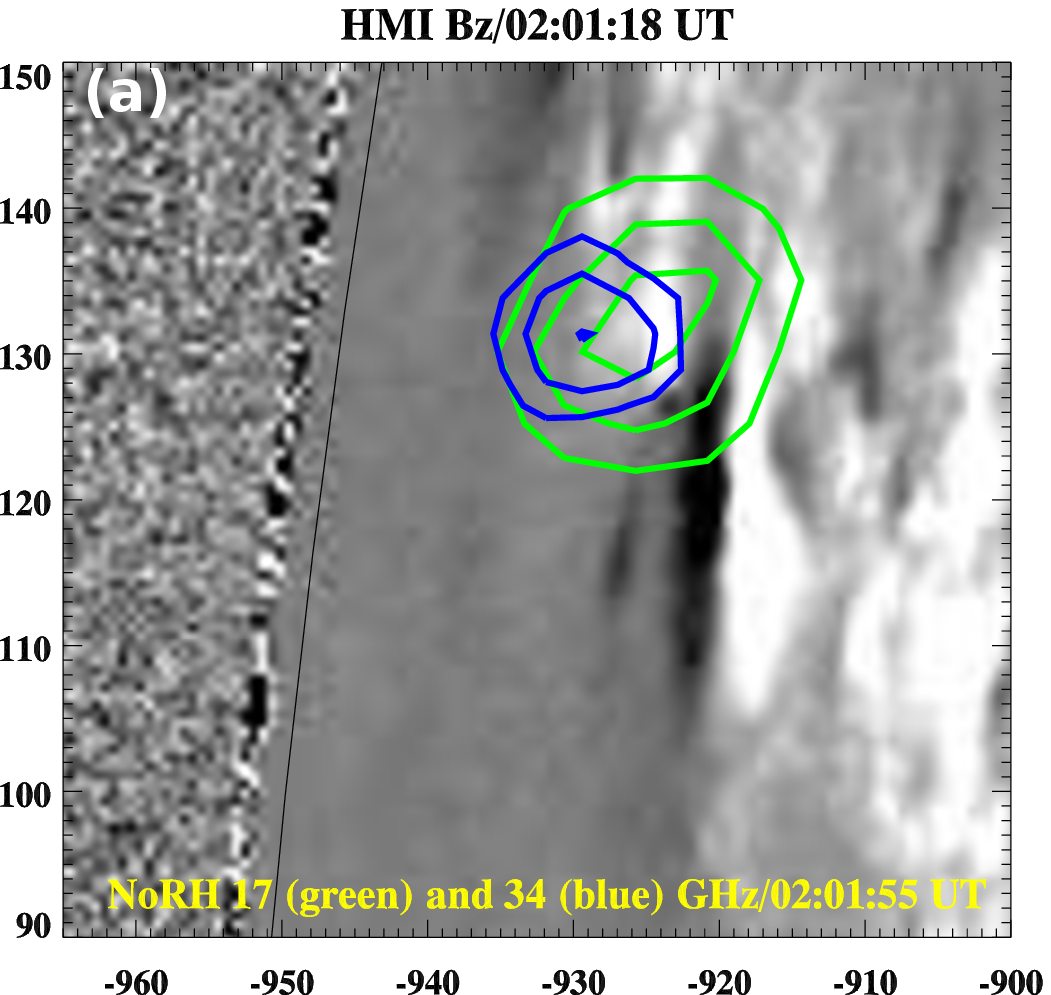}
\includegraphics[width=6.5cm]{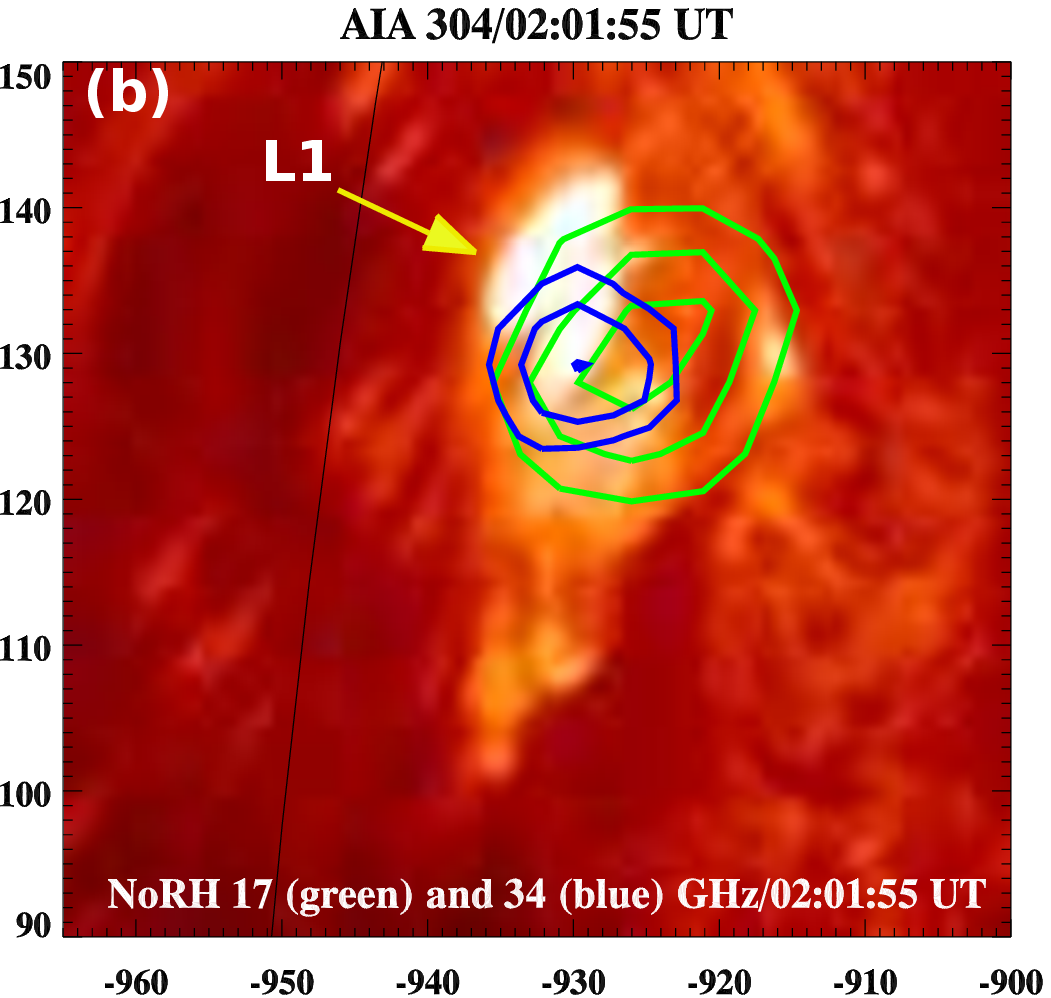}
\includegraphics[width=6.5cm]{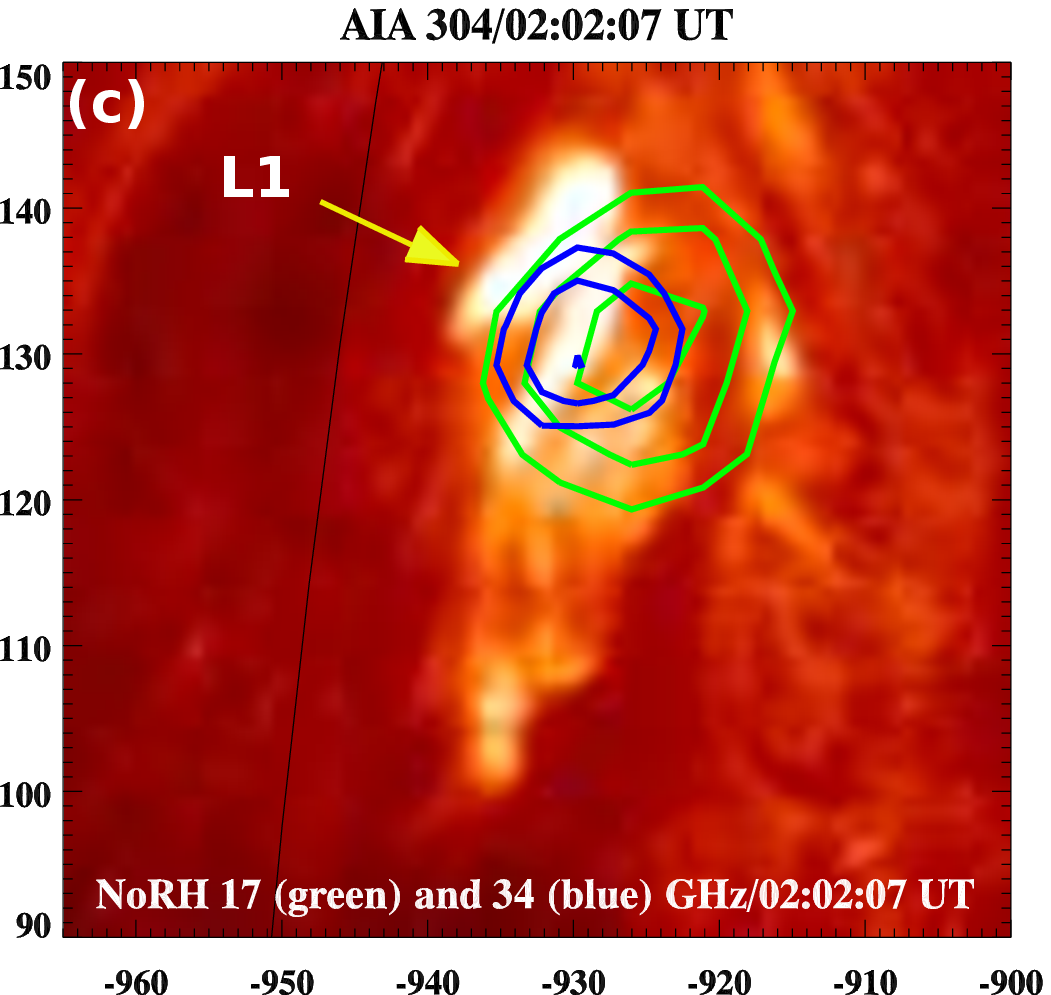}
\includegraphics[width=6.5cm]{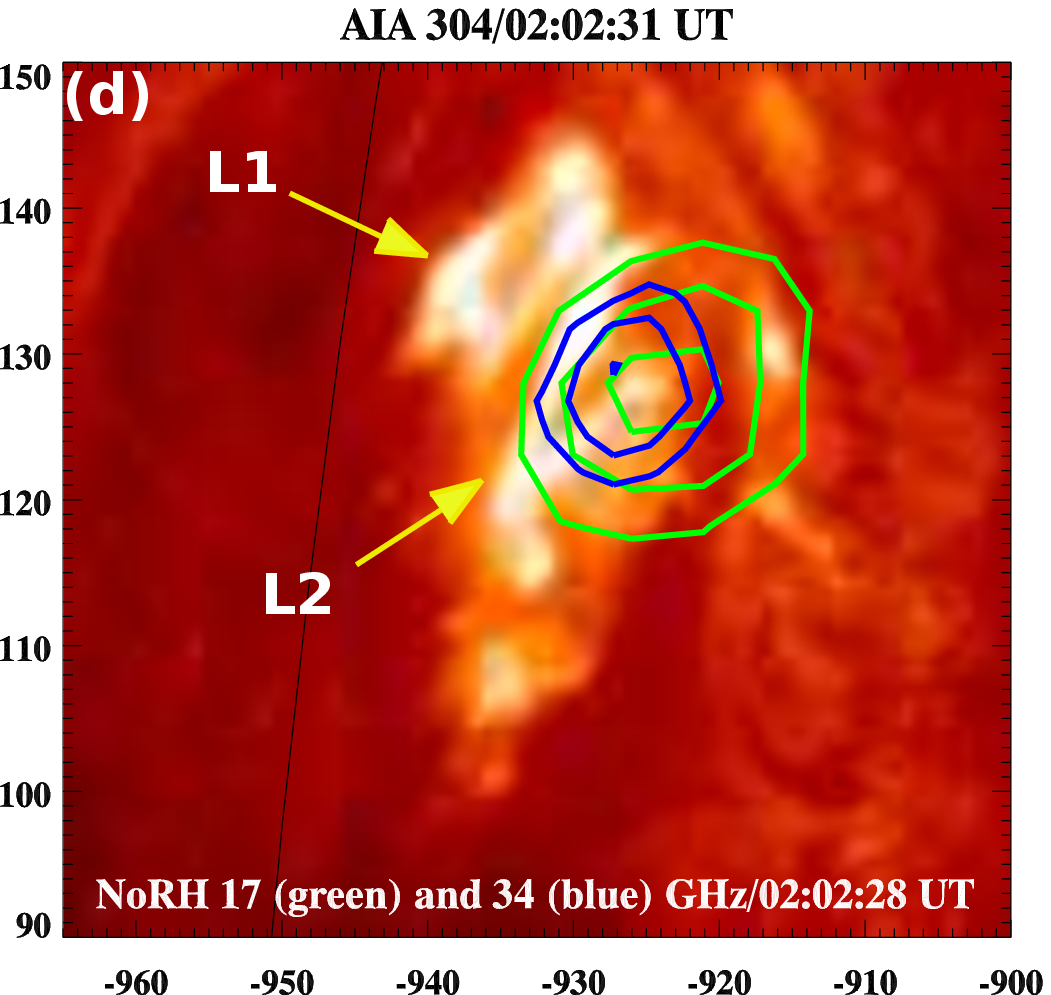}
\includegraphics[width=6.5cm]{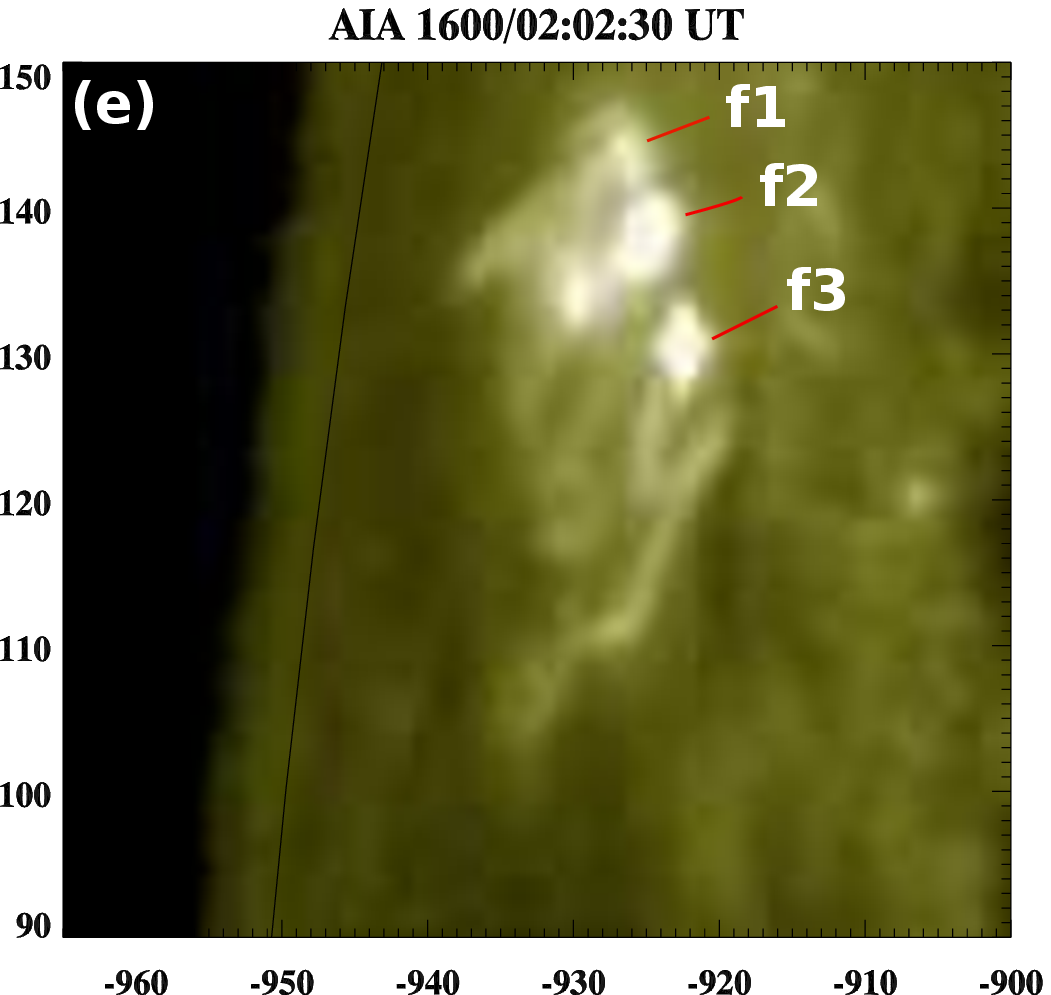}
\includegraphics[width=6.5cm]{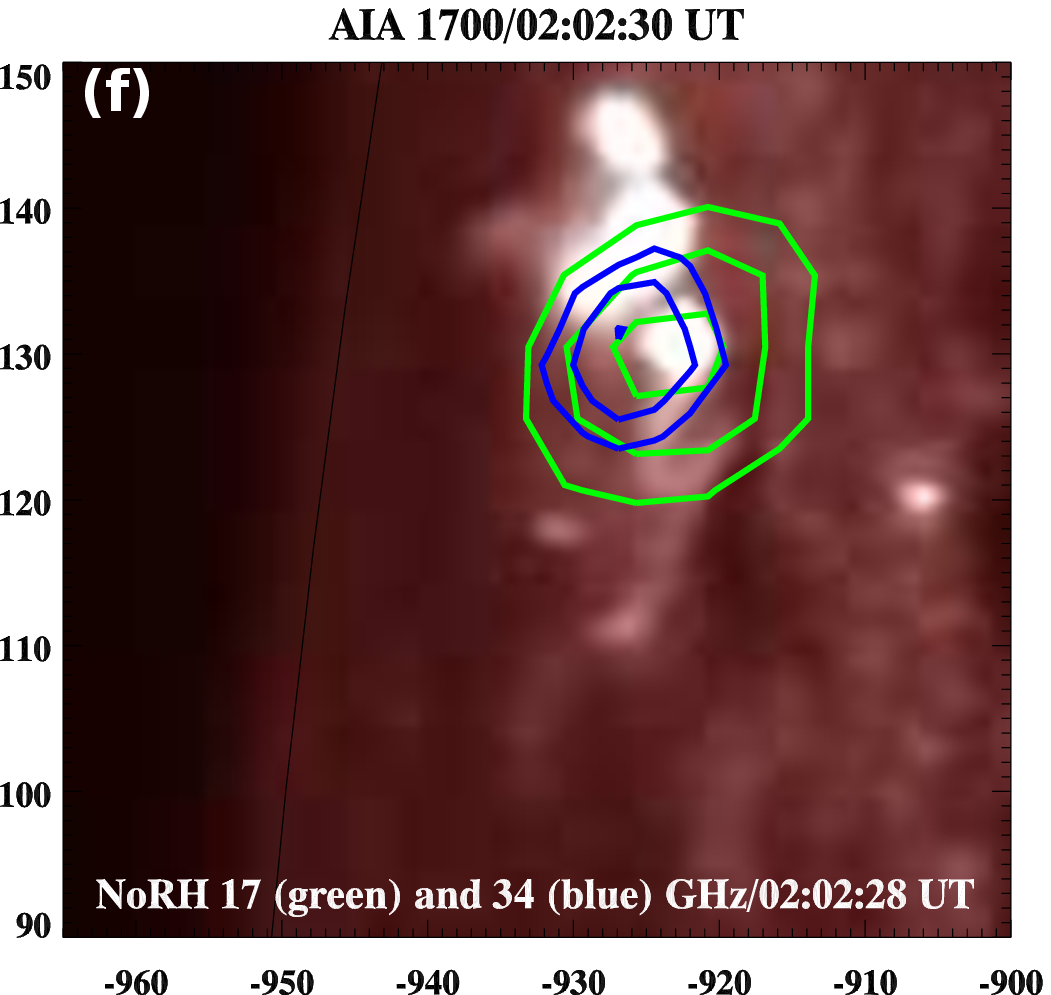}
}
\caption{HMI magnetogram (-300 to +300 Gauss) and AIA 304, 1600, and 1700 \AA~ images overlaid by NoRH 17 (green) and 34 (blue) GHz contours. The contour levels are 50\%, 70\%, and 90\% of the peak intensity. Labels f1, f2, and f3 indicate the footpoints of loops L1 and L2. f1 belongs to one of the footpoints of the rising loop L1, whereas f2 and f3 are the footpoints of loop L2.}
\label{cont}
\end{figure*}
\begin{figure*}
\centering{
\includegraphics[width=4.0cm]{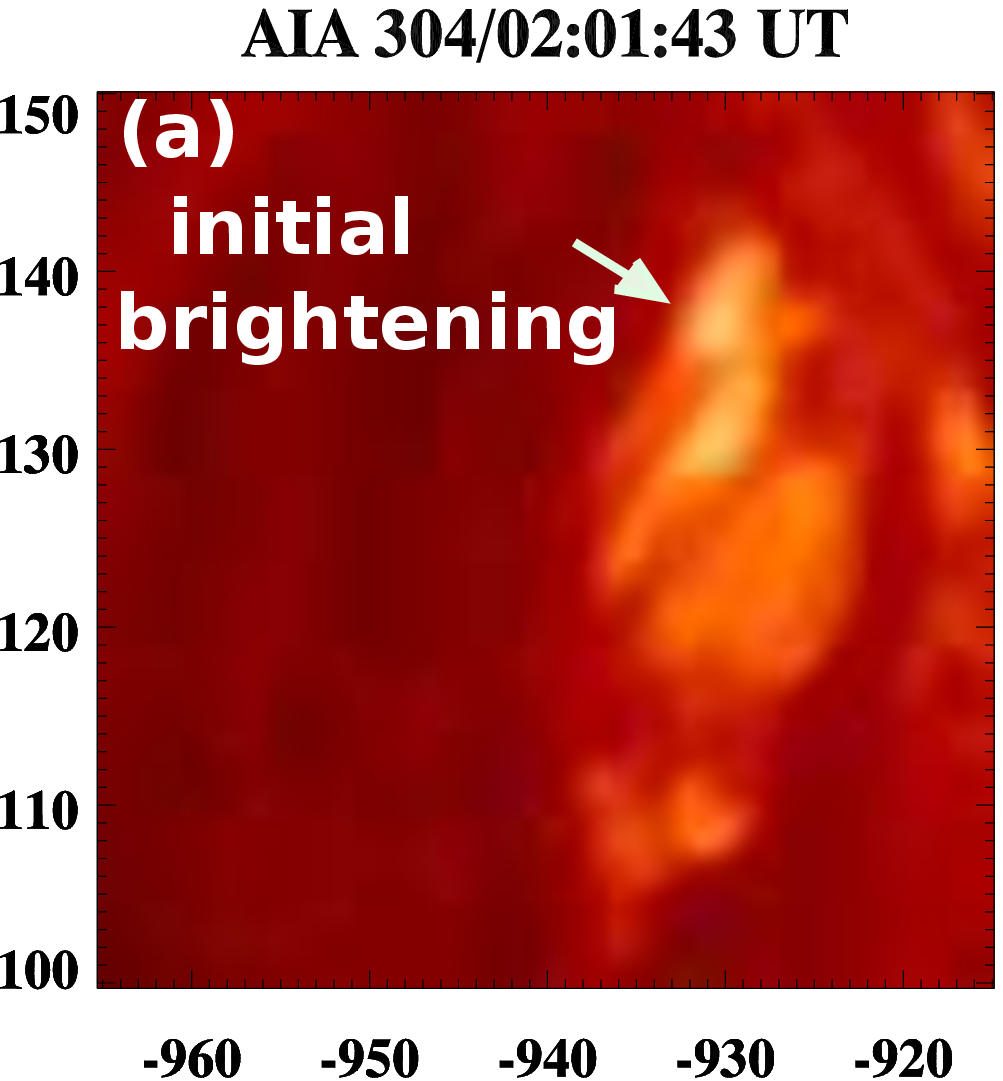}
\includegraphics[width=4.0cm]{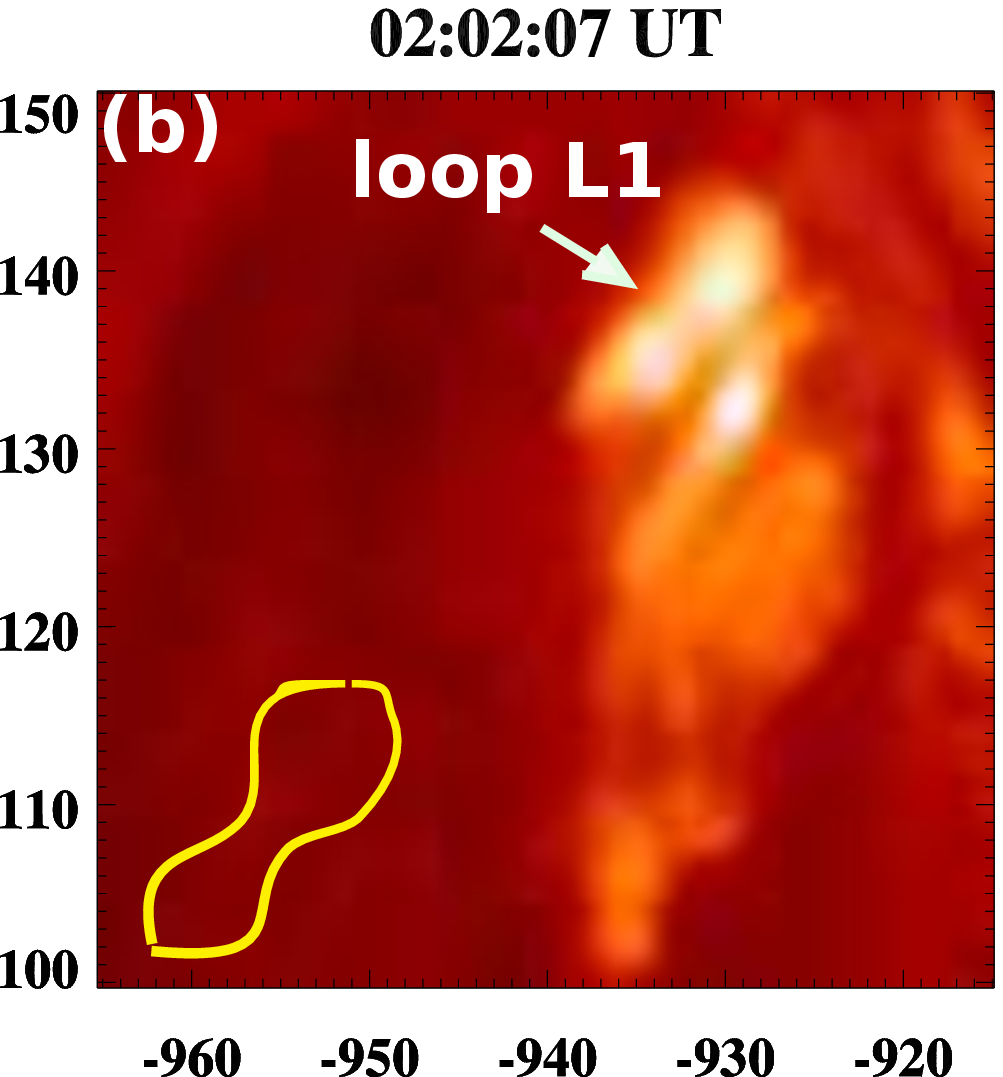}
\includegraphics[width=4.0cm]{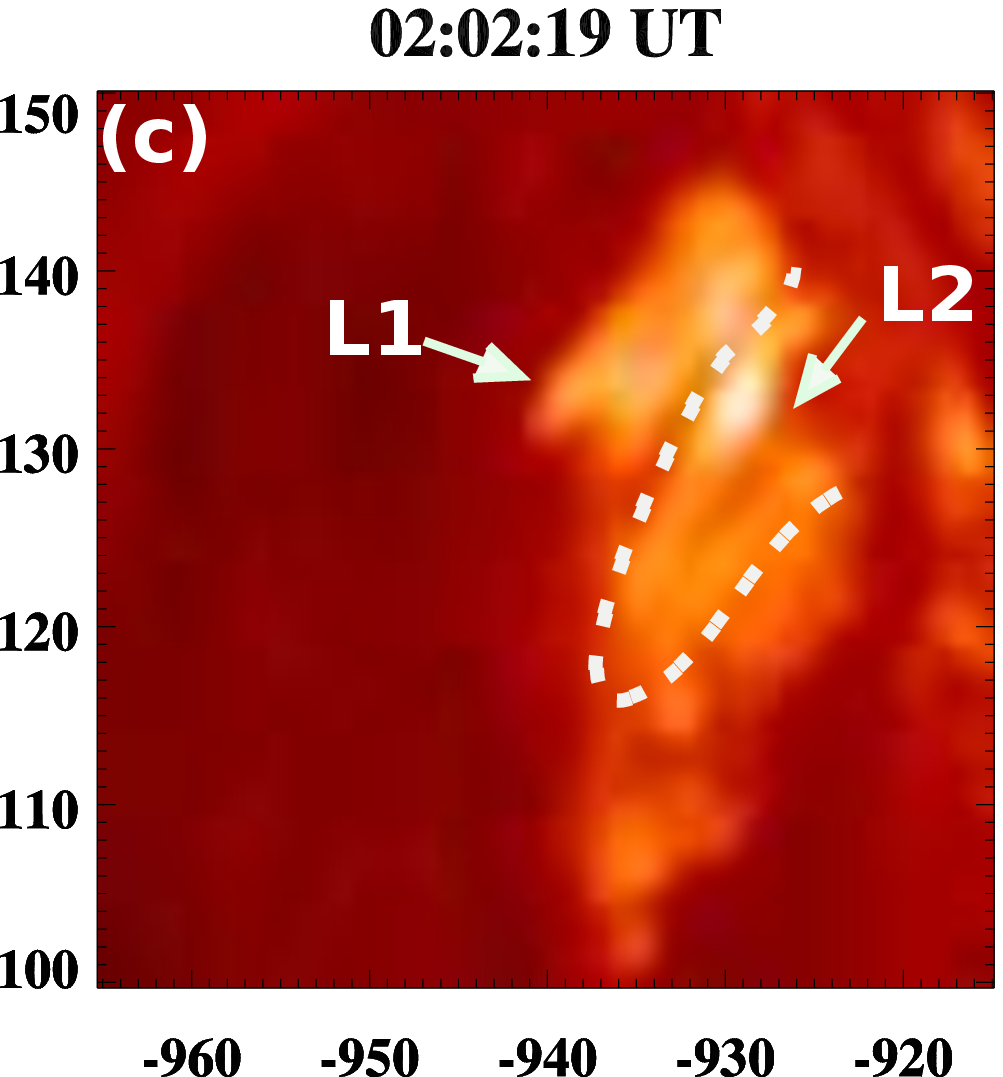}

\includegraphics[width=4.0cm]{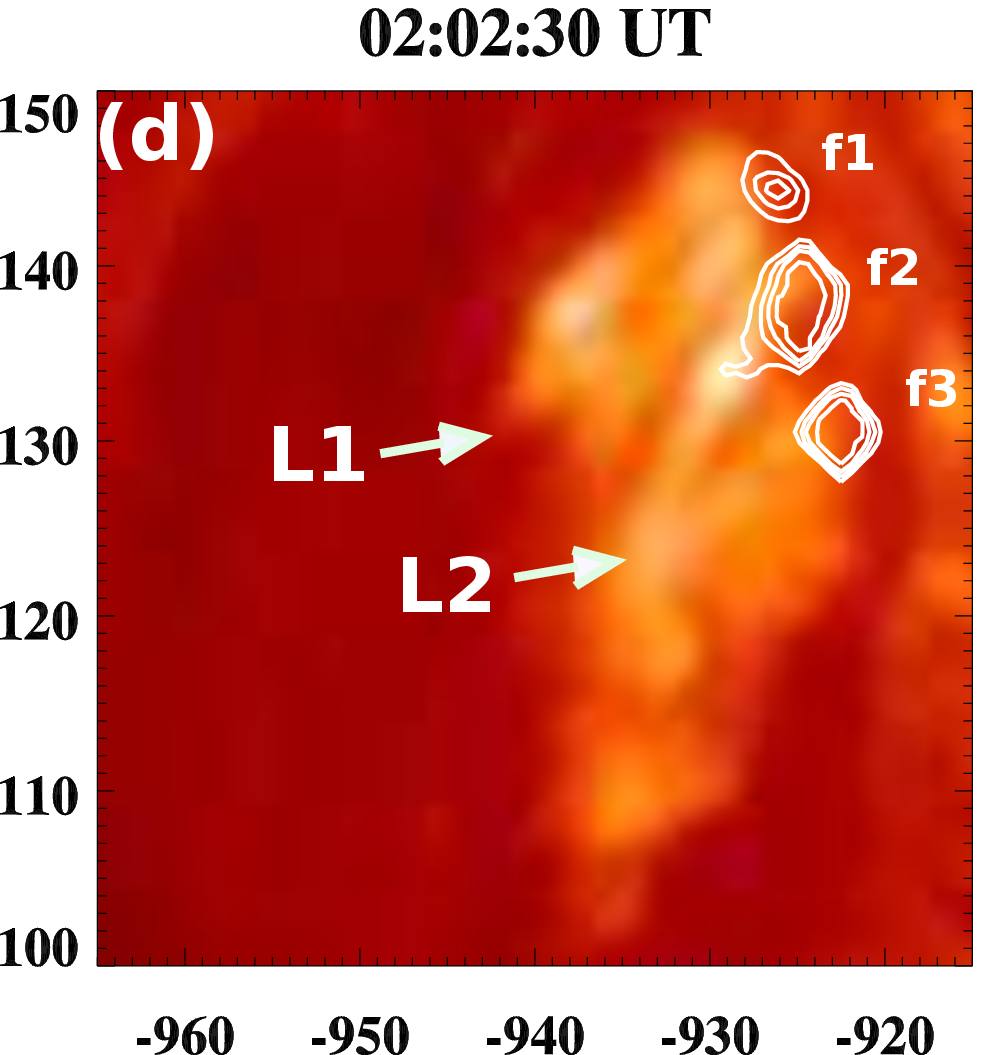}
\includegraphics[width=4.0cm]{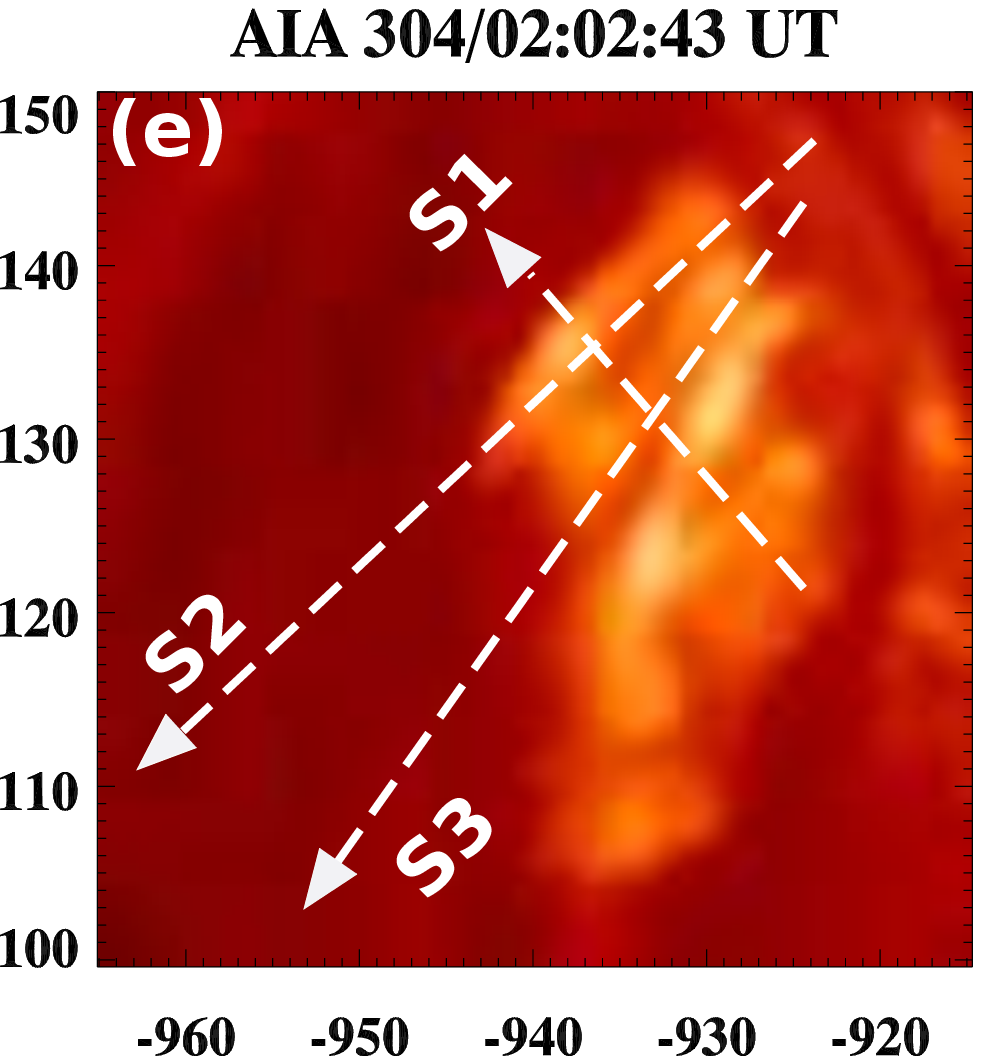}
\includegraphics[width=4.0cm]{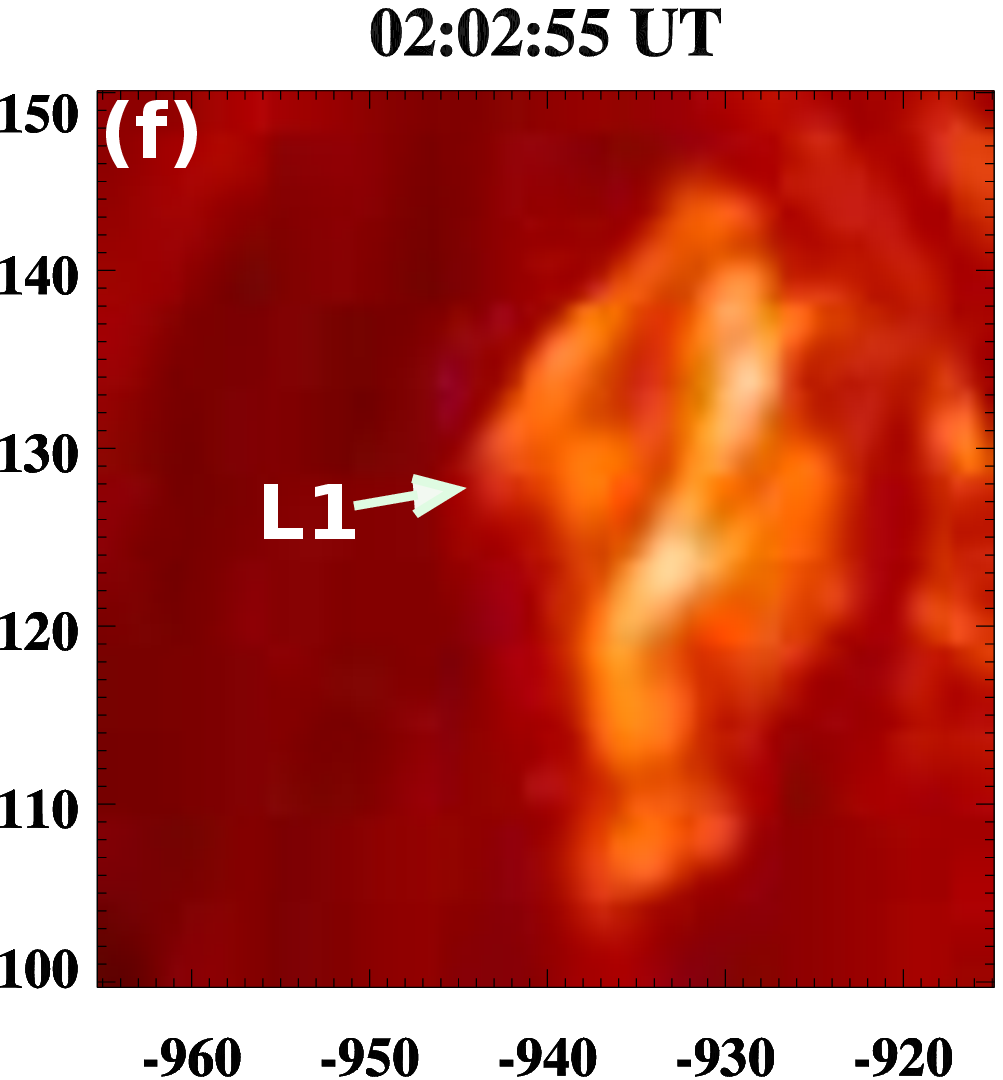}

\includegraphics[width=4.0cm]{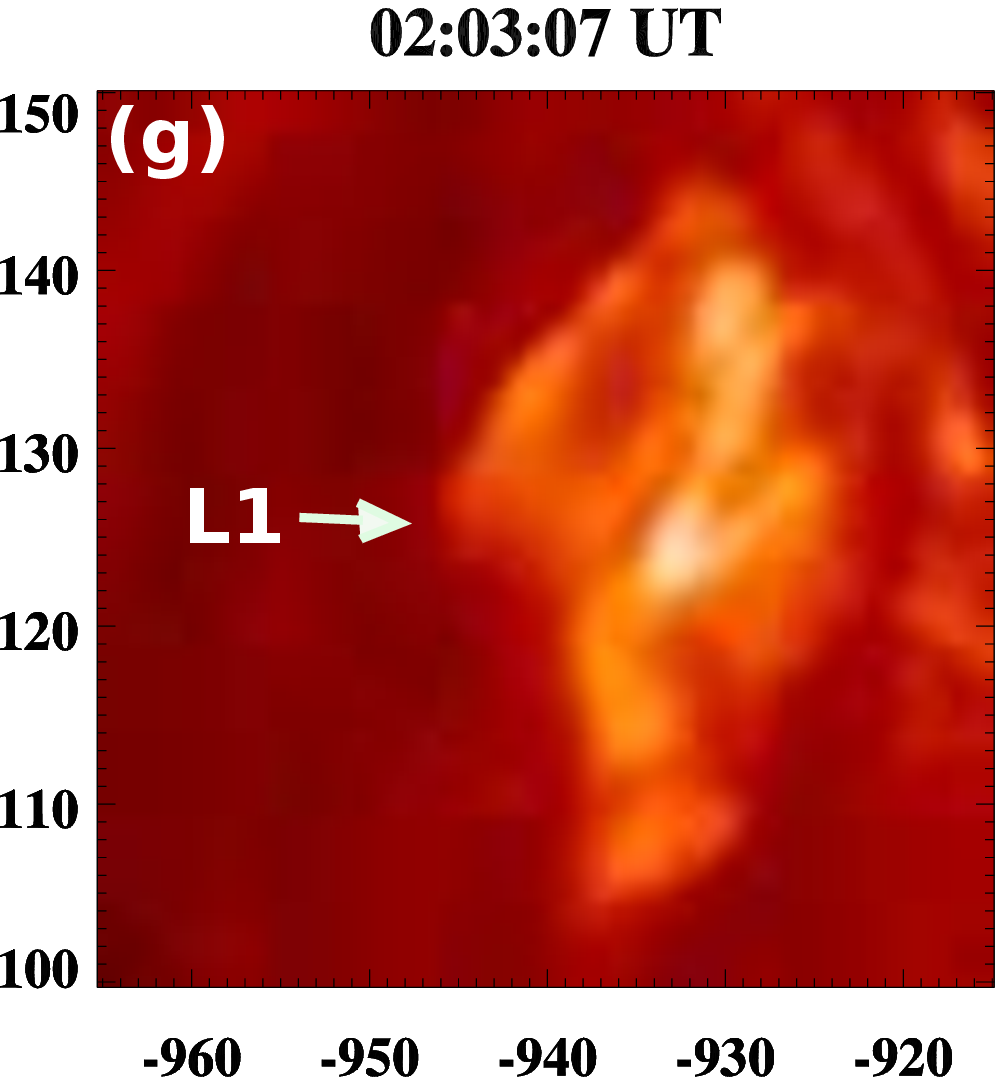}
\includegraphics[width=4.0cm]{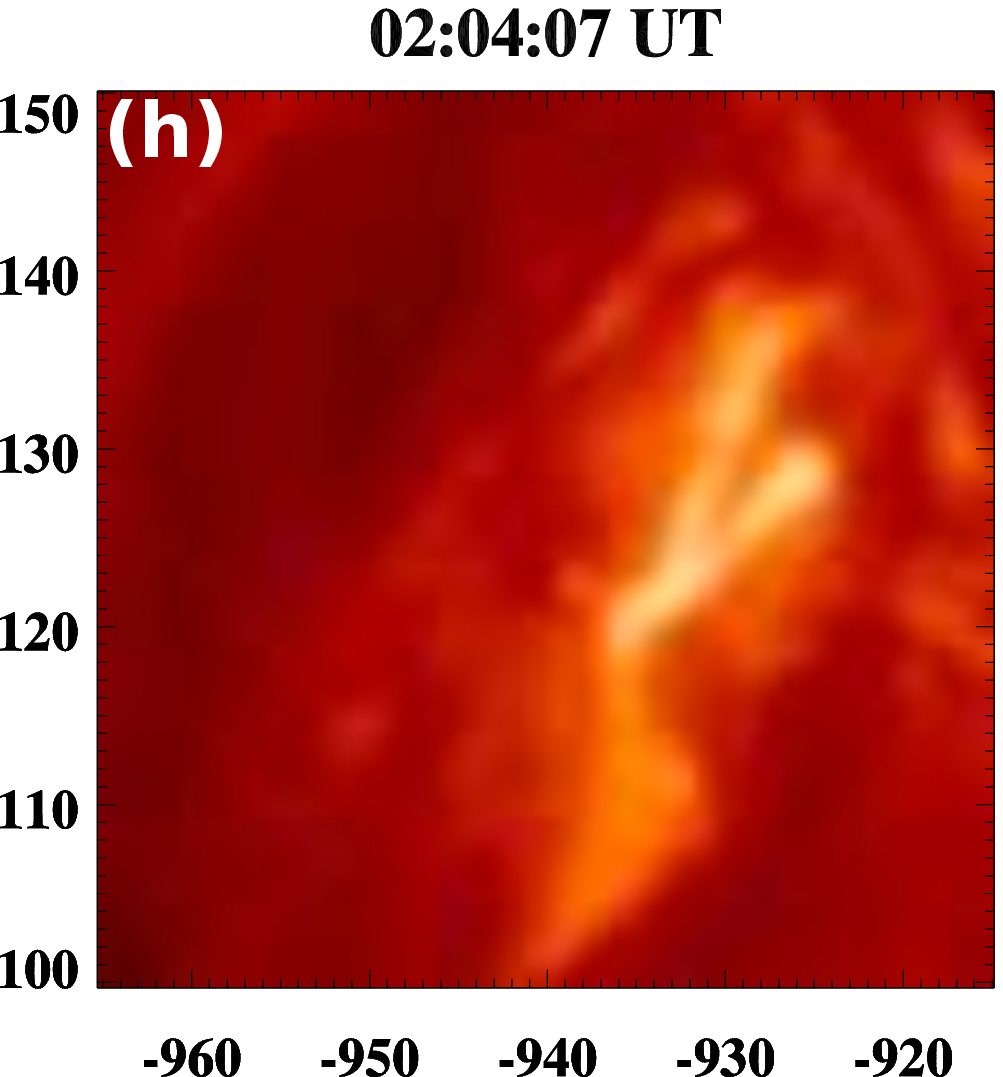}
\includegraphics[width=4.0cm]{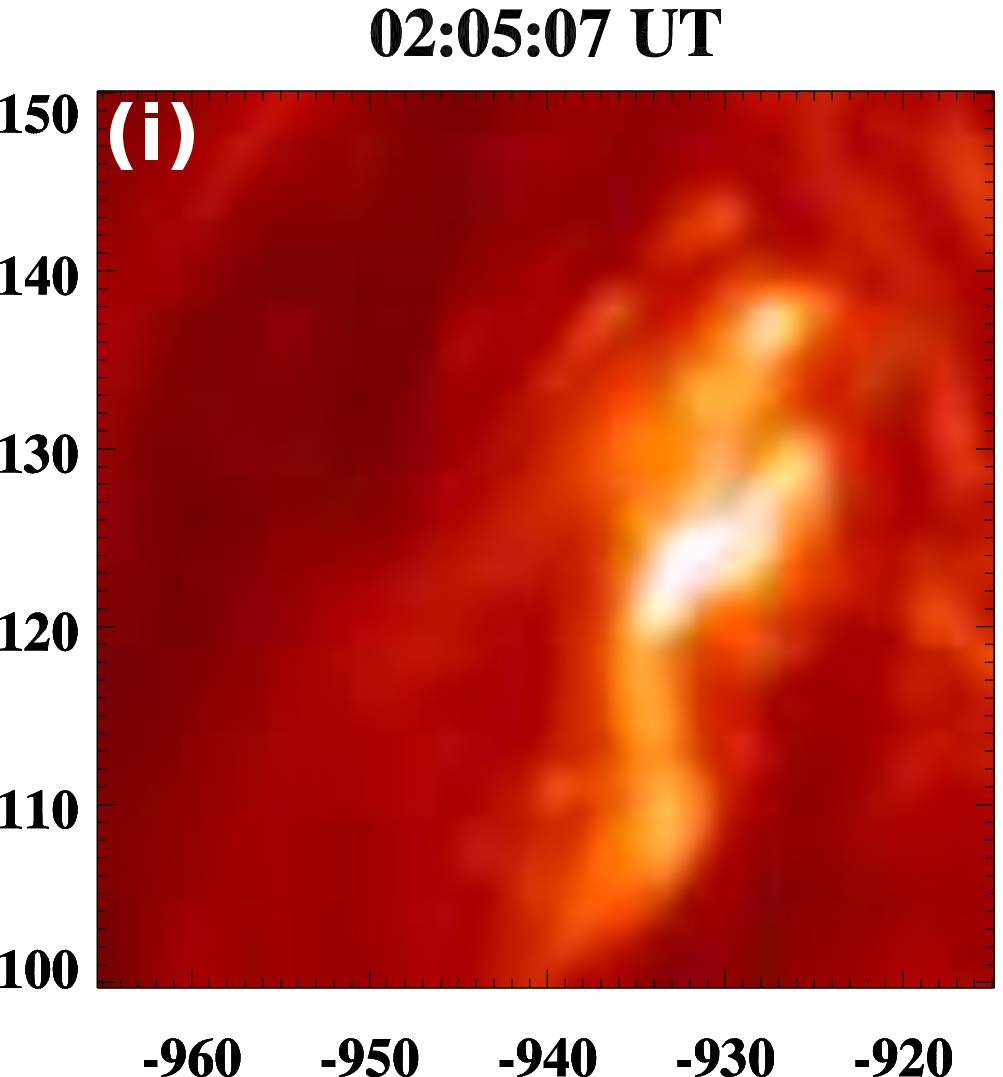}

\includegraphics[width=4.0cm]{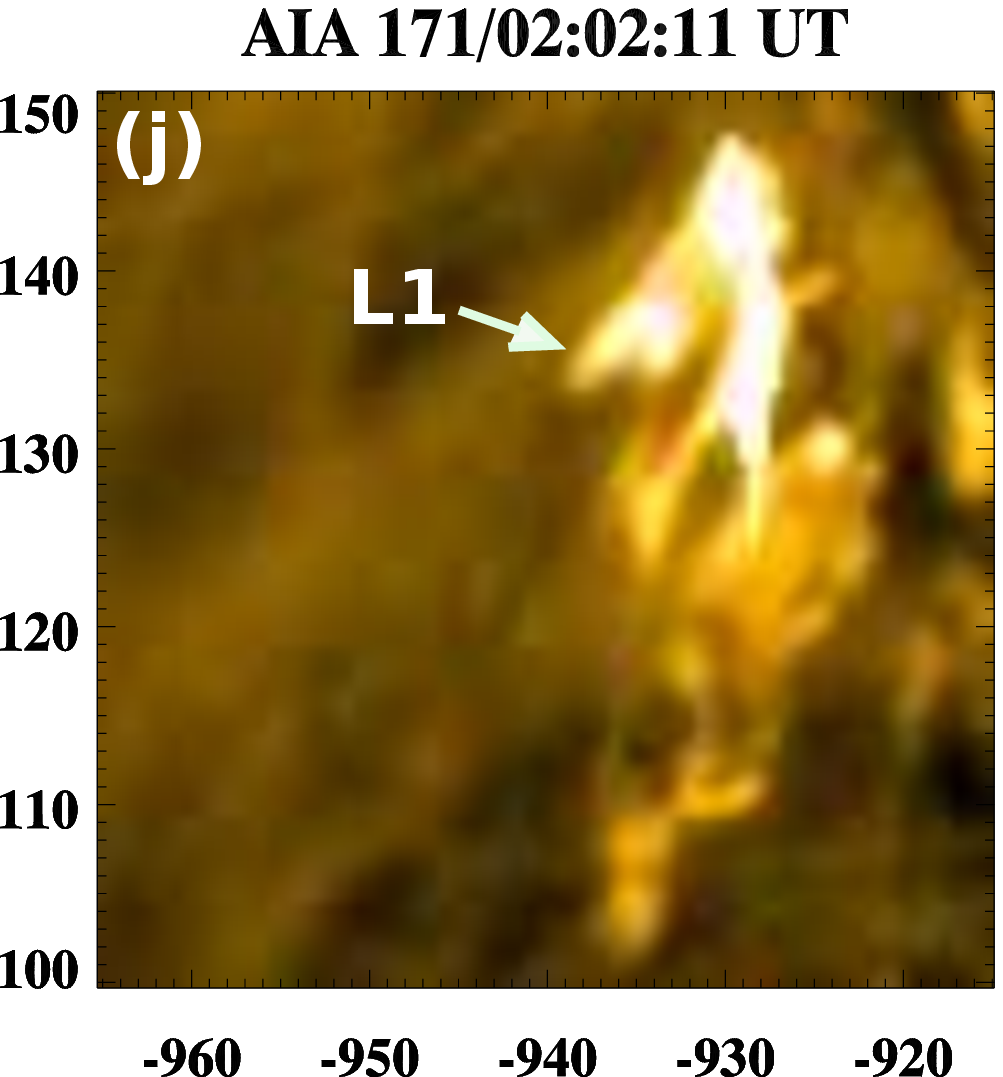}
\includegraphics[width=4.0cm]{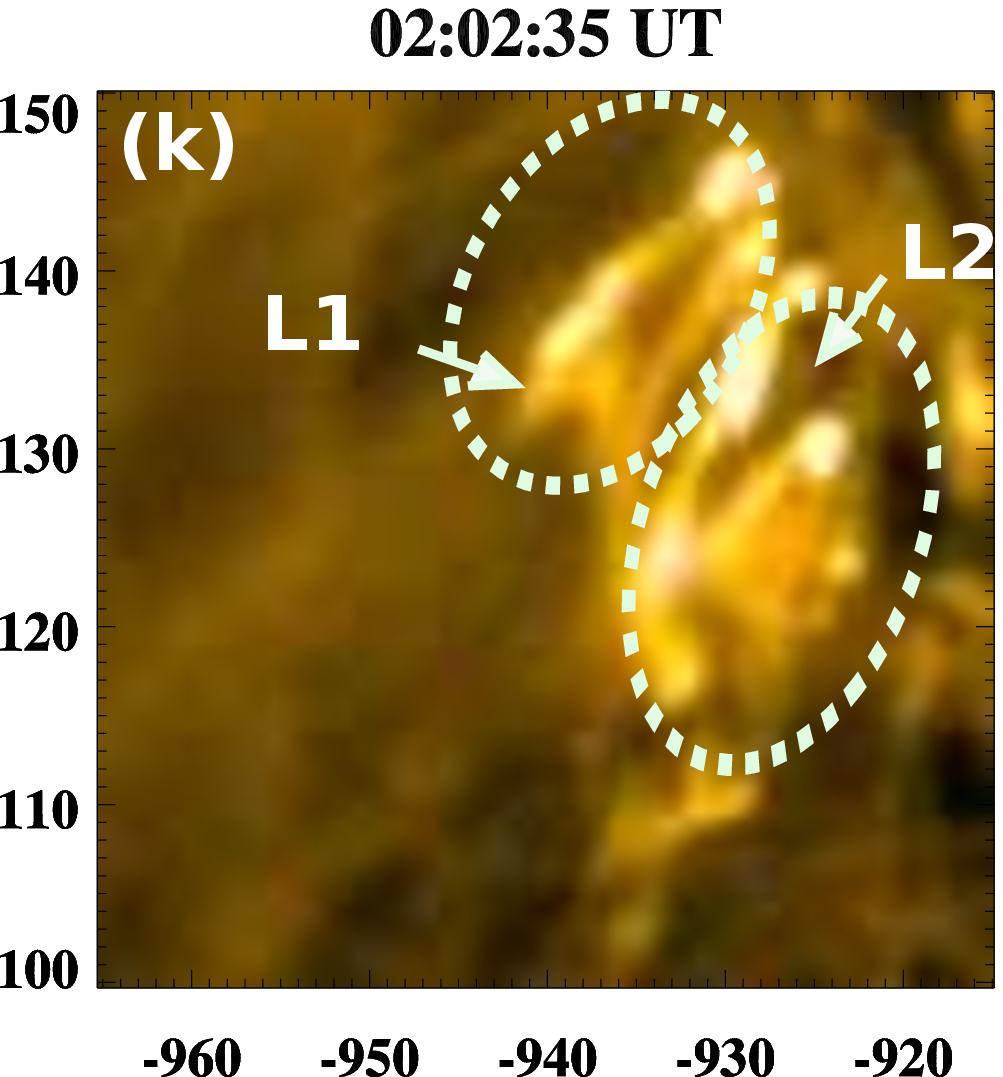}
\includegraphics[width=4.0cm]{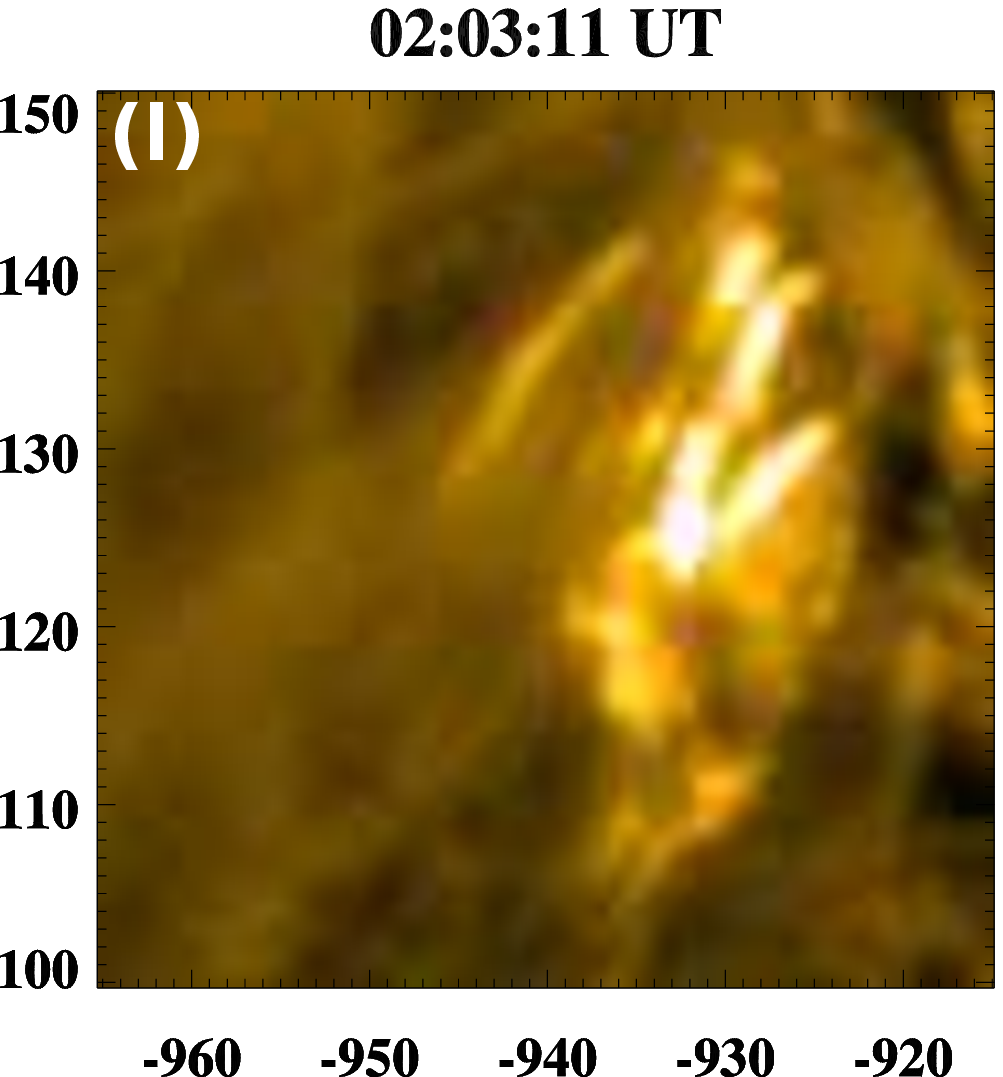}
}
\caption{AIA 304 and 171 \AA~ image sequence during the flare showing a thick loop, L1, moving radially outward and the appearance of a cusp-shaped loop, L2, nearby. The yellow sketch at the corner of panel (b) indicates the initial structure of the bright loop L1. The 13-s QPP was observed during the rising motion of L1 ($\sim$02:01:45-02:02:45~UT). Labels S1, S2, and S3 show the slice cuts across and along the loops, taken for the time-distance plots. AIA 1700 \AA~ contours (white) are overlaid in the AIA 304 \AA~ image (panel (d)) to show the emission from the footpoints (f1, f2, f3) of the interacting loops. The contour levels are 35\%, 55\%, 75\%, 95\% of the peak intensity
{\it(An animation of this figure is available)}.}
\label{aia304_171}
\end{figure*}

\begin{figure*}
\centering{
\includegraphics[width=8.5cm]{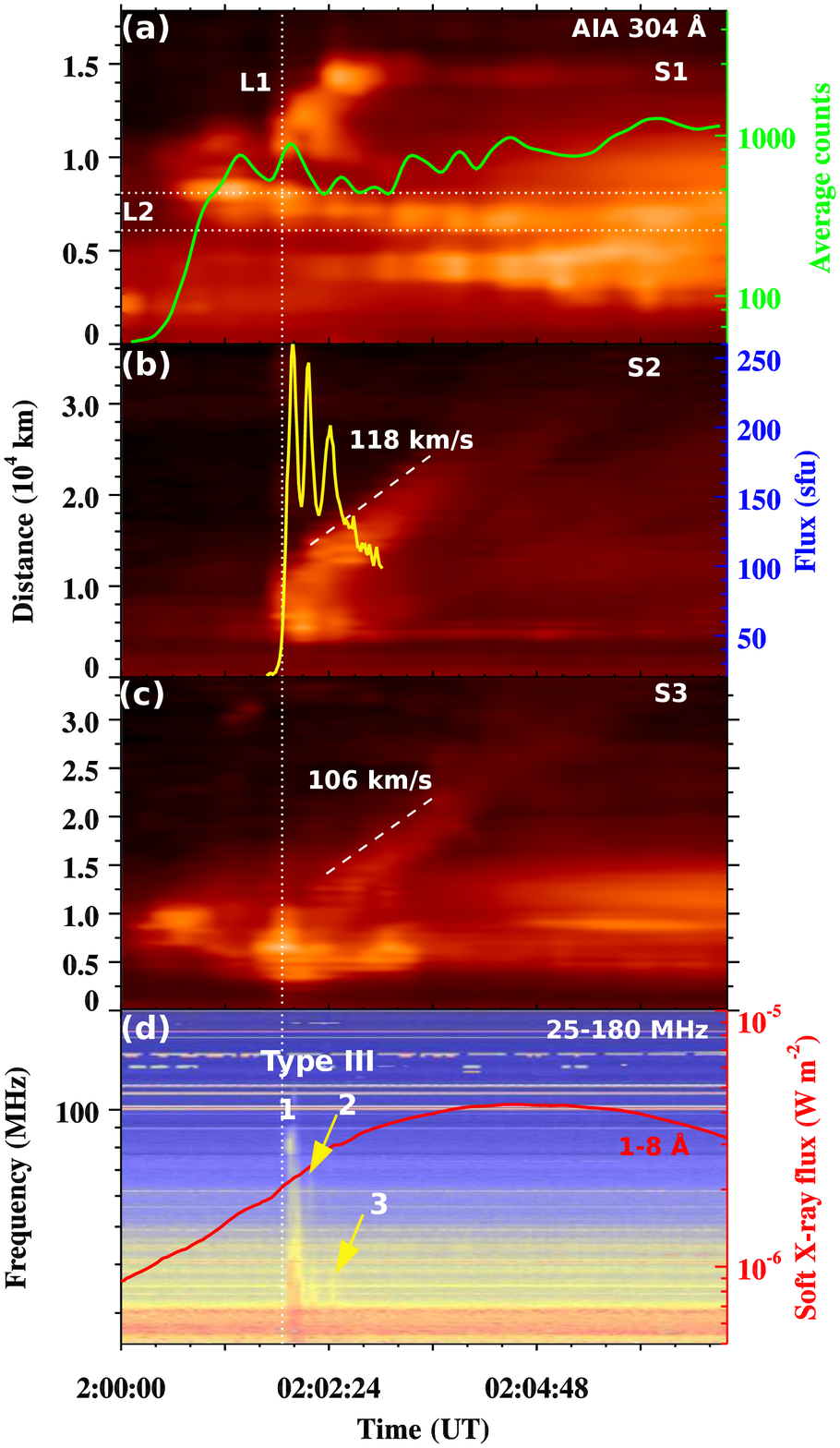}
\includegraphics[width=7.5cm]{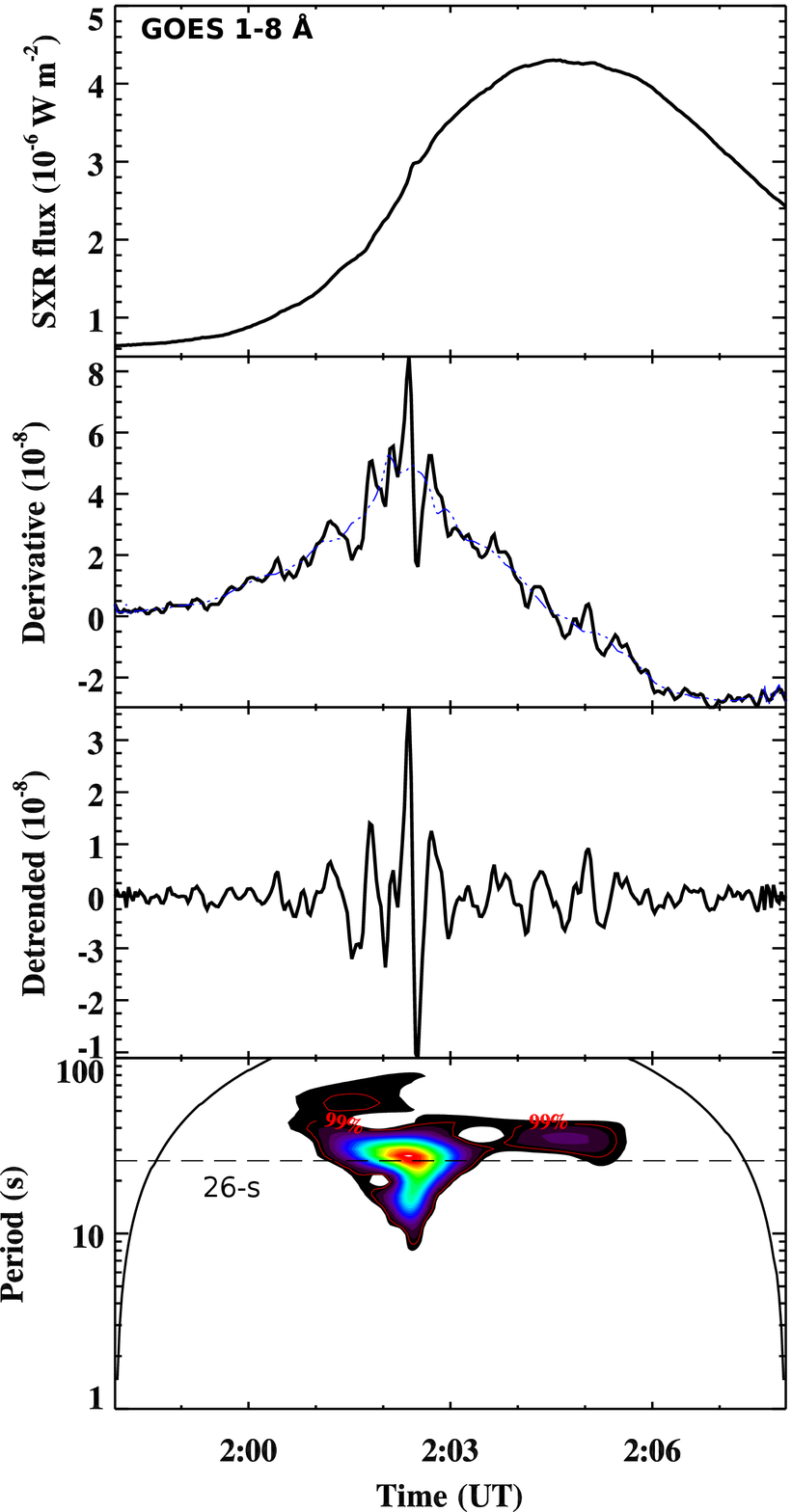}
}
\caption{{\it Left:} Time-distance plots of the AIA 304 \AA~ intensity along slices S1, S2, and S3. The green curve is the average smoothed counts extracted between the two horizontal dotted lines in the time-distance plot of AIA 304 \AA~ intensity. The yellow curve shows the NoRH 34~GHz flux in sfu. The dynamic radio spectrum (25--180 MHz) showing quasi-periodic type III radio bursts (marked by 1, 2, and 3) was observed at the Learmonth solar observatory. A vertical dotted line indicates the onset of QPP associated with a rising loop in AIA 304 \AA~ channel. {\it Right:} GOES soft X-ray flux (2-s cadence) in 1--8 \AA~ channel, its derivative, smoothed/detrended, and wavelet power spectrum. The dash-dotted curve indicated a 60-s smoothing of the signal.}
\label{st_304}
\end{figure*}
\begin{figure*}
\centering{
\includegraphics[width=5.5cm]{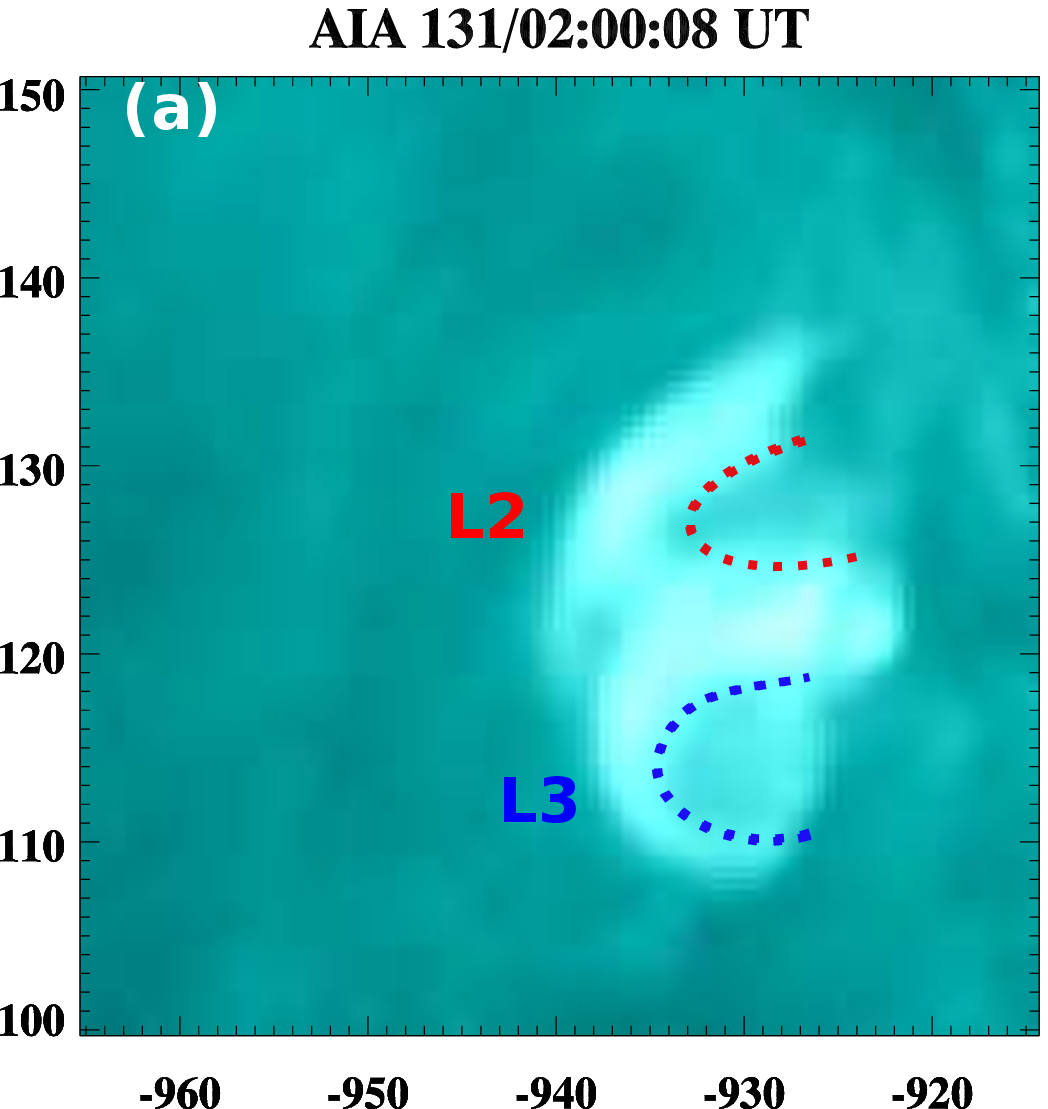}
\includegraphics[width=5.5cm]{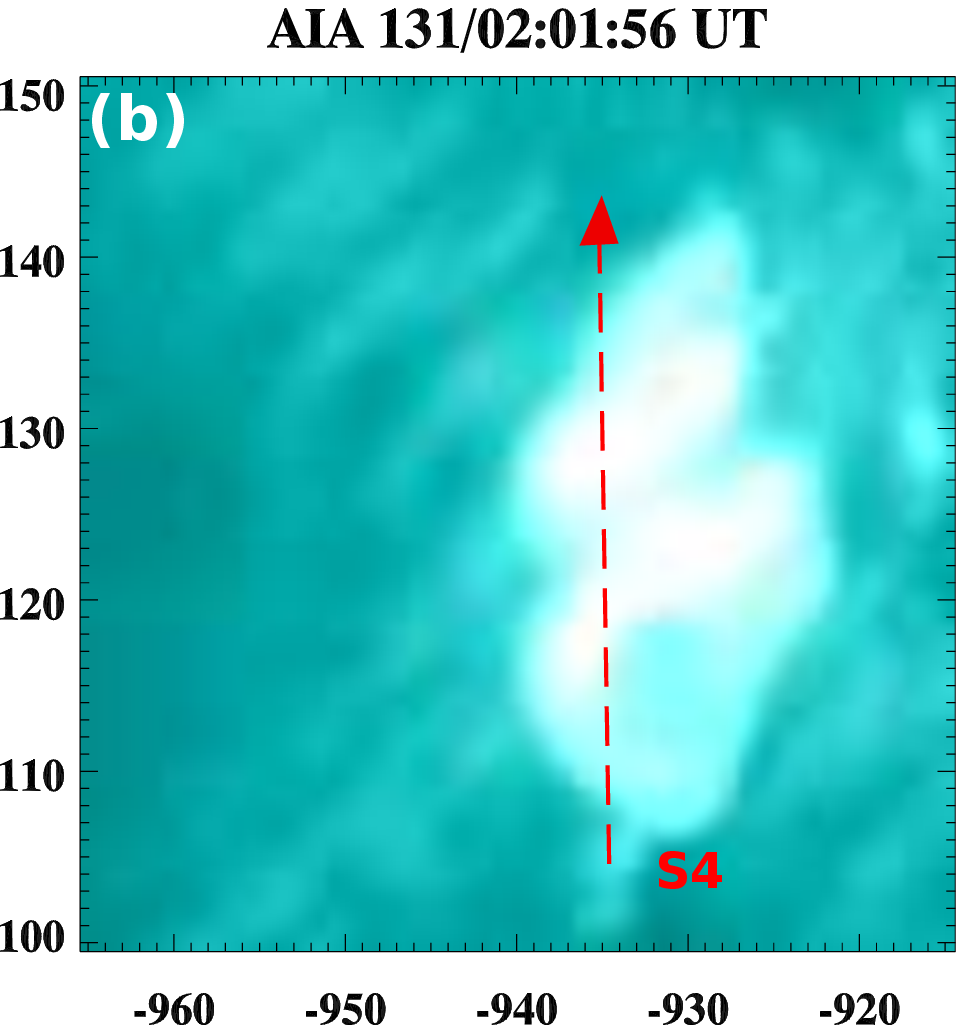}

\includegraphics[width=5.5cm]{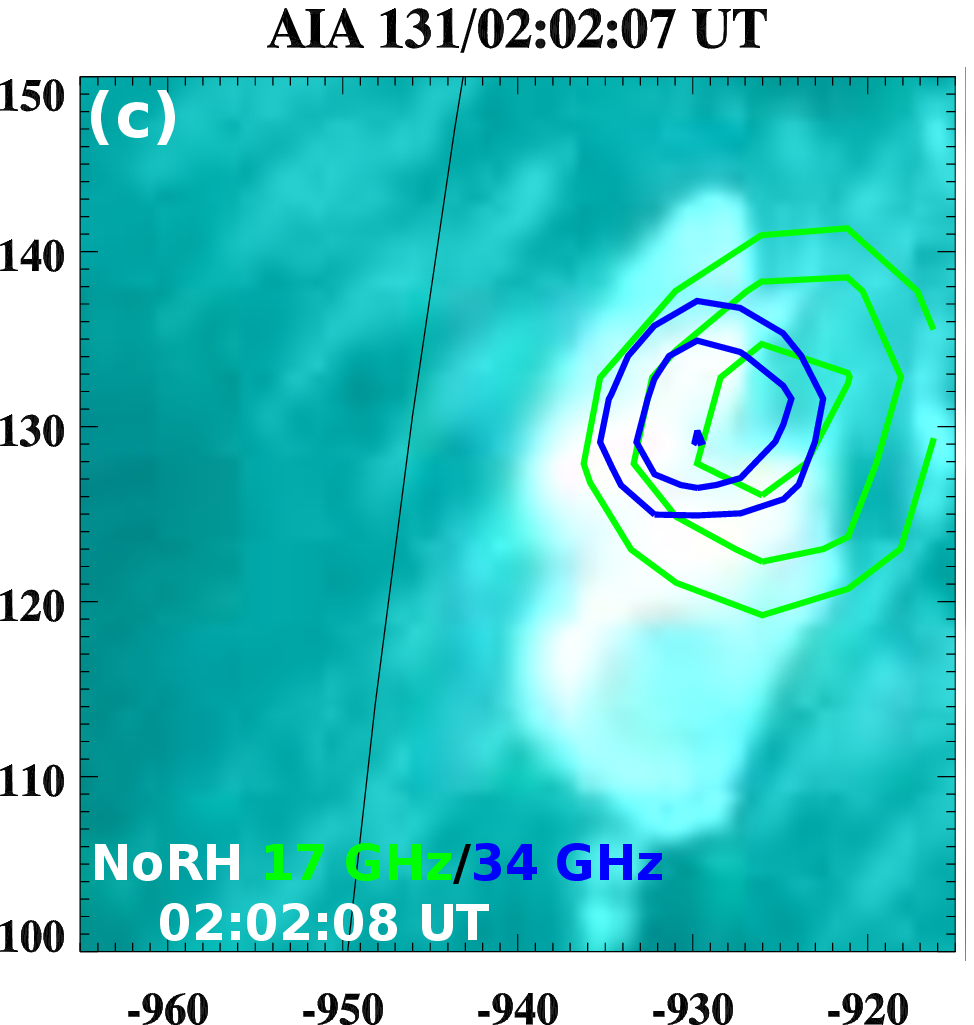}
\includegraphics[width=5.5cm]{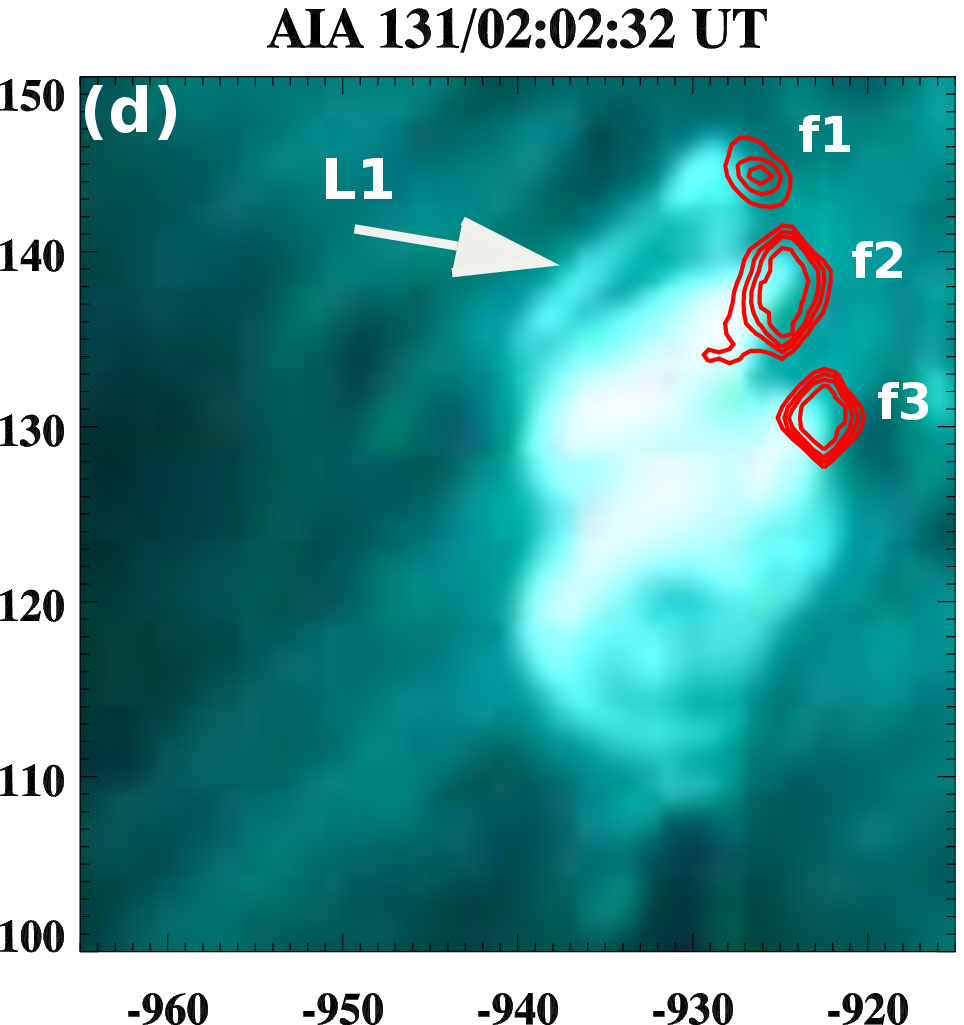}

\includegraphics[width=5.5cm]{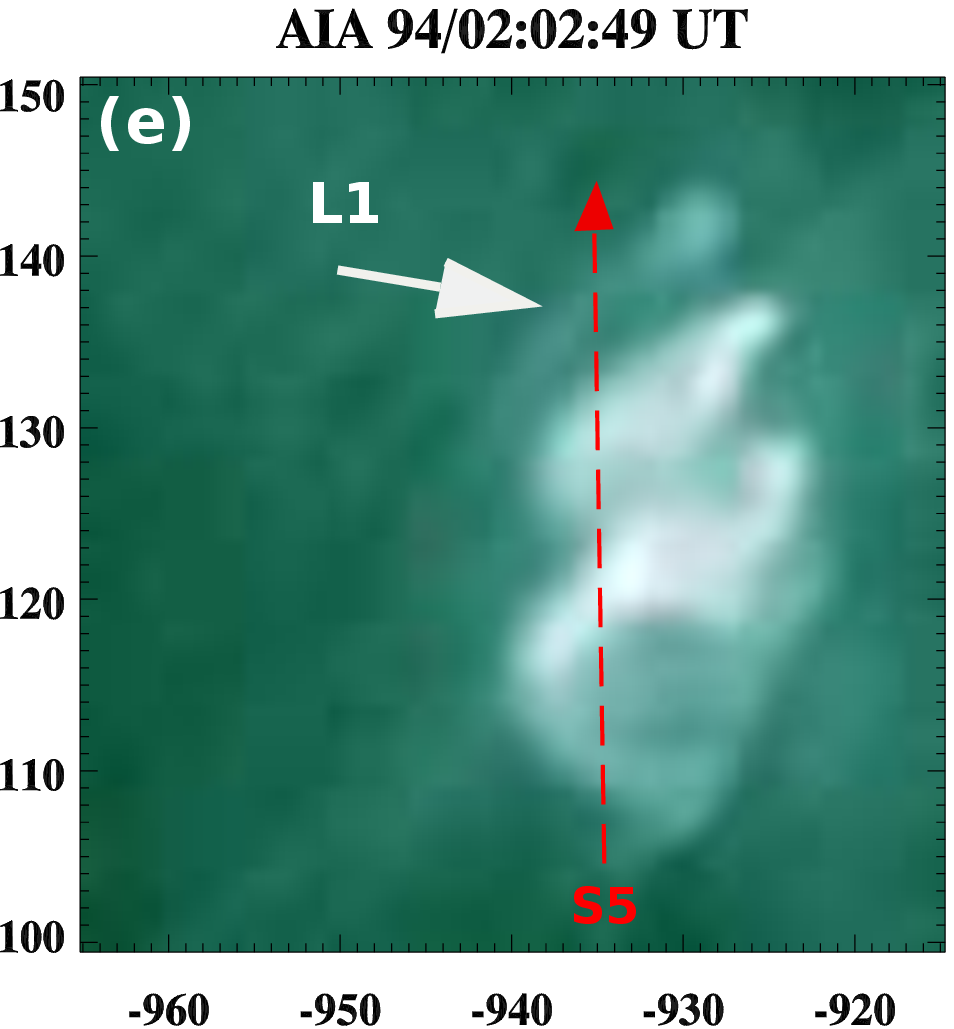}
\includegraphics[width=5.5cm]{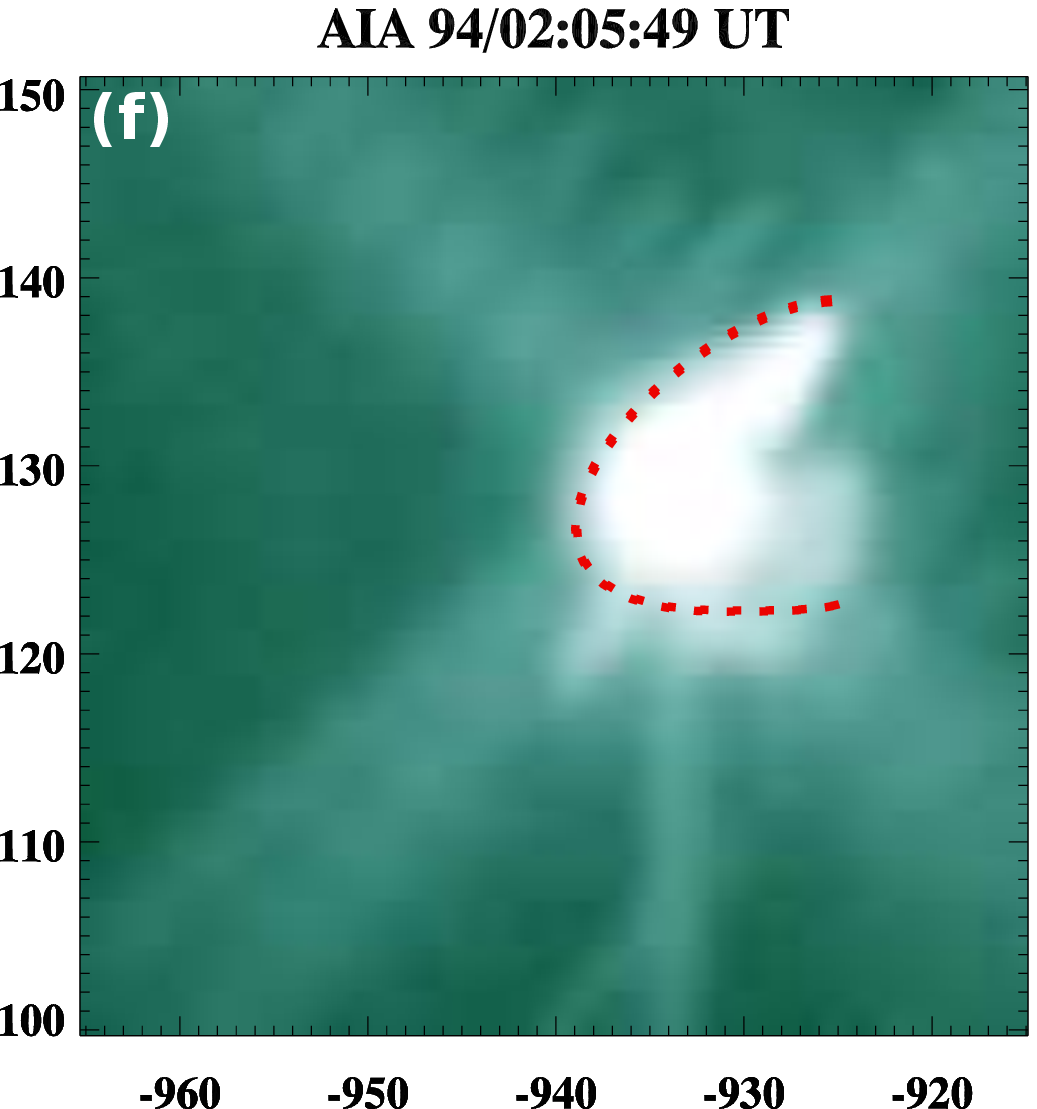}

}
\caption{AIA 131 and 94 \AA~ images during the flare showing interacting the loops (L2 and L3). AIA 131 \AA~ image (panel (c)) is overlaid by NoRH 17 (green) and 34 (blue) GHz contours. The contour levels are 50\%, 70\%, and 90\% of the peak intensity. S4 and S5 are the slice cuts used to construct the time-distance intensity plots. Arrows indicate the rising loop, L1. AIA 1700 \AA~ contours (red, 02:02:30 UT) are overlaid on the AIA 131 \AA~ image (panel (d)) to show the emission from the footpoints (f1, f2 and f3) of the loops. The contour levels are 35\%, 55\%, 75\%, 95\% of the peak intensity {\it(An  animation of this figure is available)}.}
\label{aia131_94}
\end{figure*}
\begin{figure}
\centering{
\includegraphics[width=9cm]{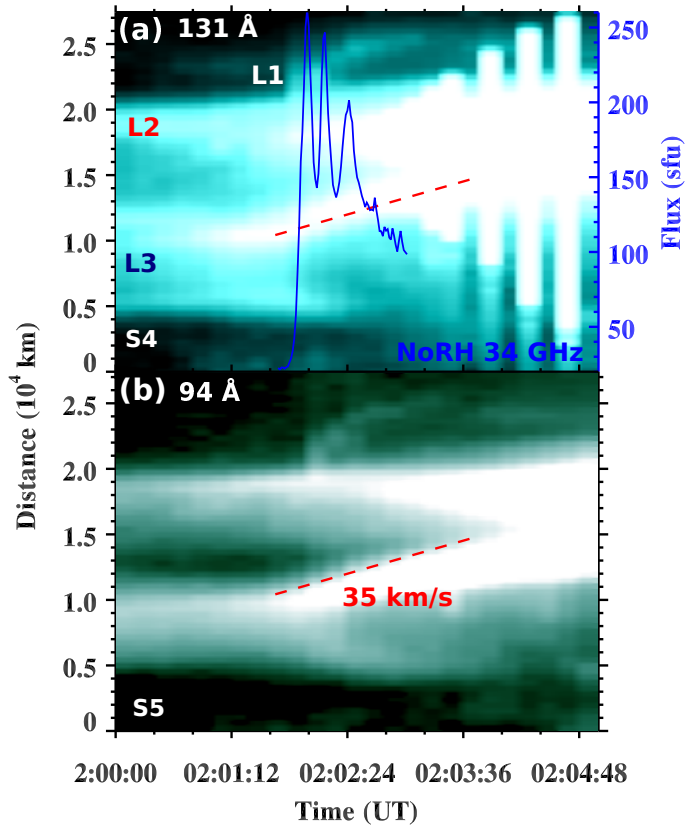}
}
\caption{Time-distance intensity plots along the slices S4 and S5 using AIA 131 and 94 \AA~ images. The blue curve is the NoRH 34 GHz flux in sfu.}
\label{st_131}
\end{figure}
\begin{figure*}
\centering{
\includegraphics[width=7cm]{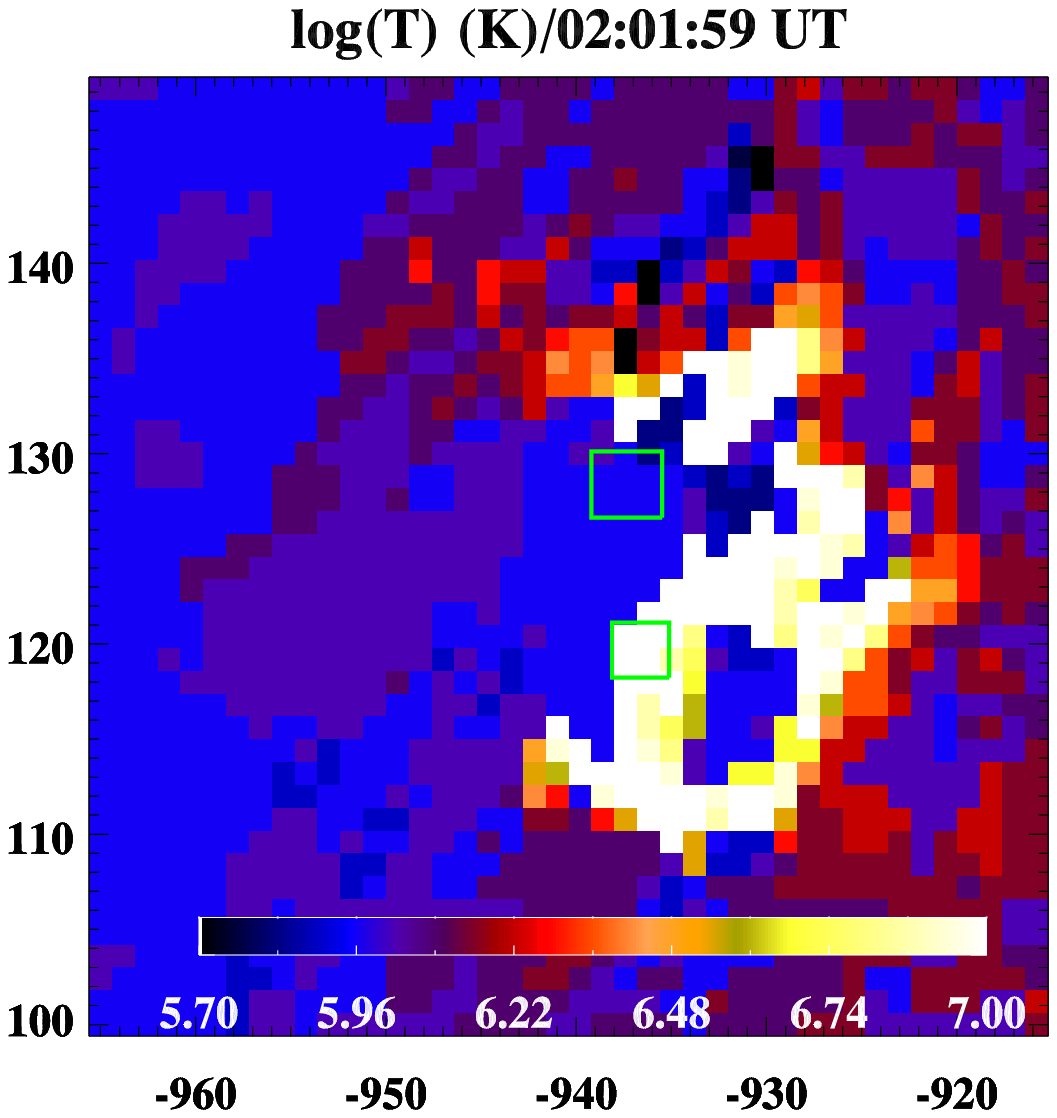}
\includegraphics[width=7cm]{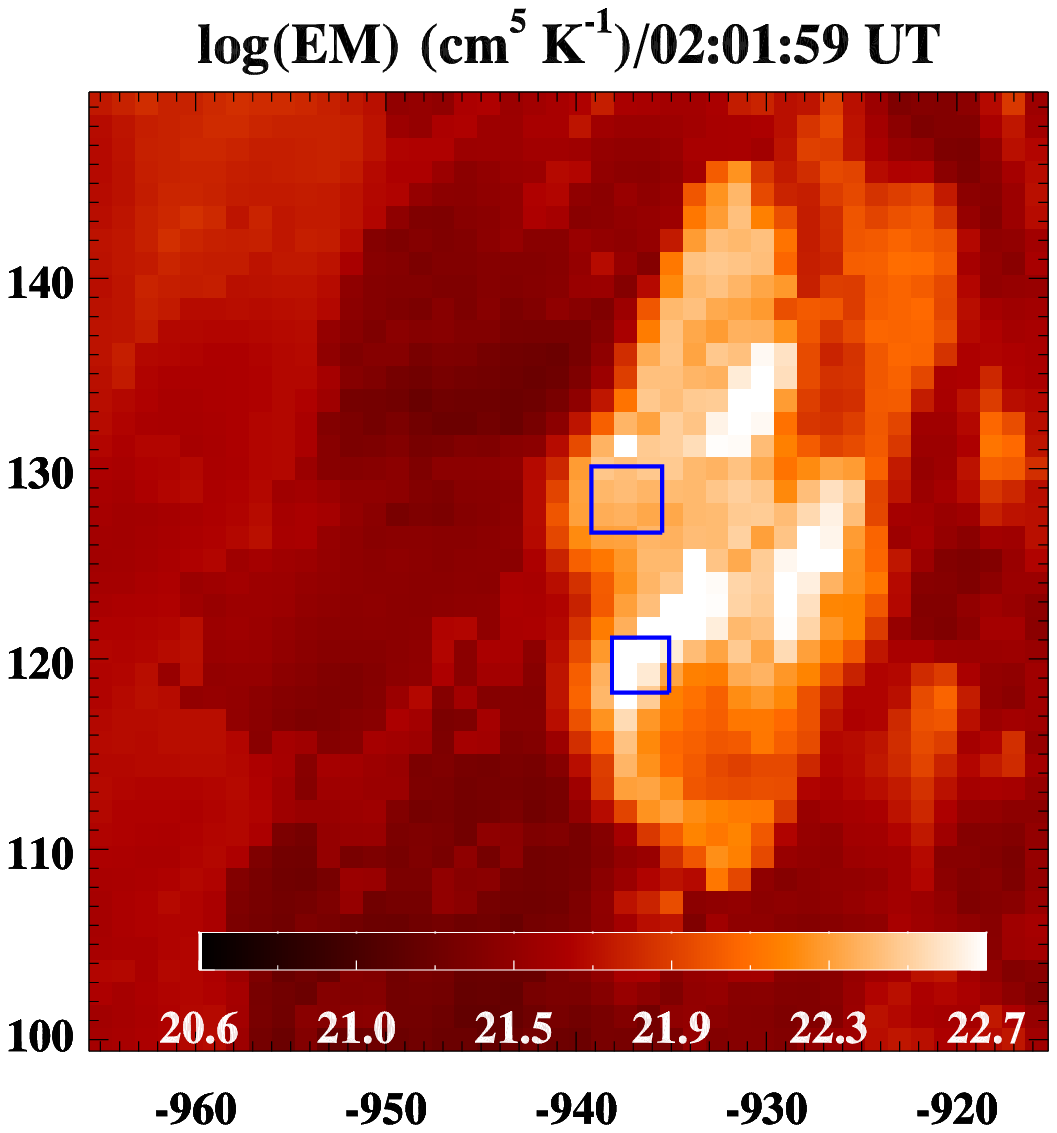}
}
\caption{Peak temperature ($T$) and emission measure (EM) maps of the flare site derived from the near simultaneous images in six AIA channels. The green and blue boxes are used to estimate the mean $T$ and EM of the loops L2 and L3.}
\label{dem}
\end{figure*}
\begin{figure*}
\centering{
\includegraphics[width=5.3cm]{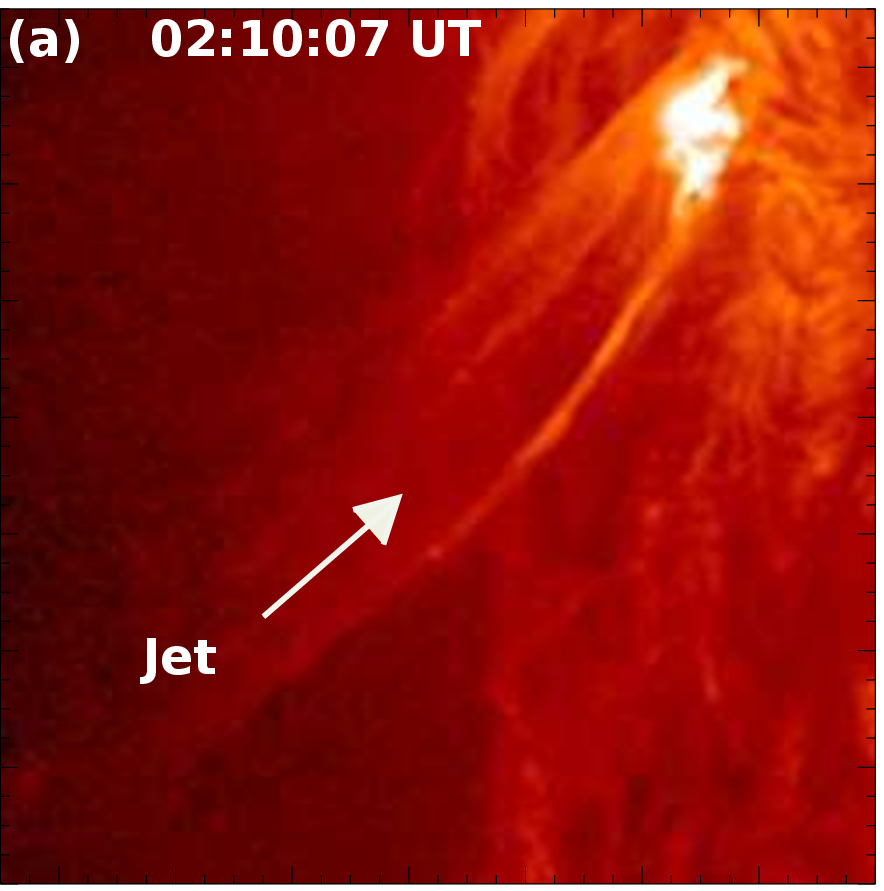}
\includegraphics[width=5.4cm]{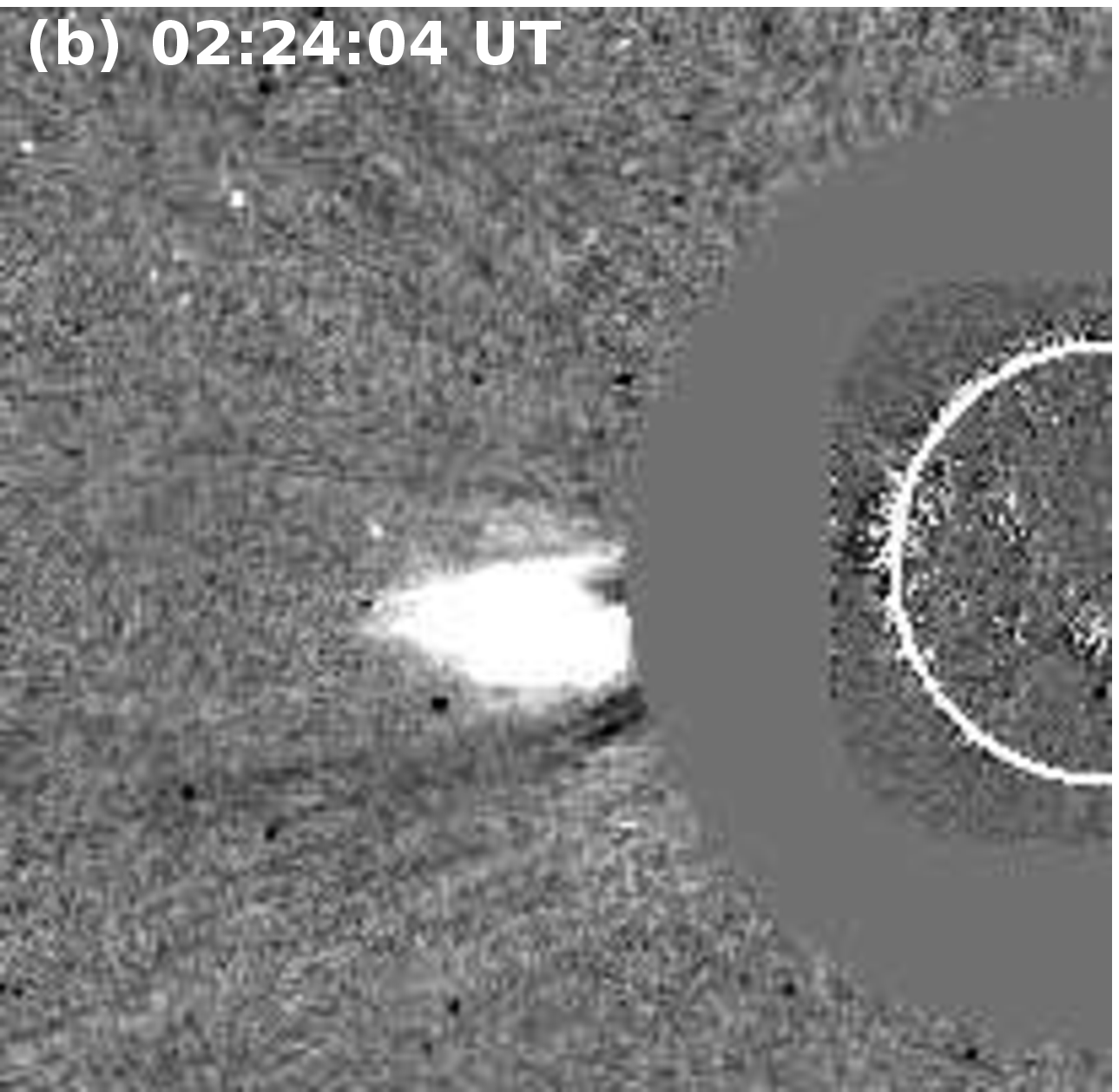}
\includegraphics[width=5.4cm]{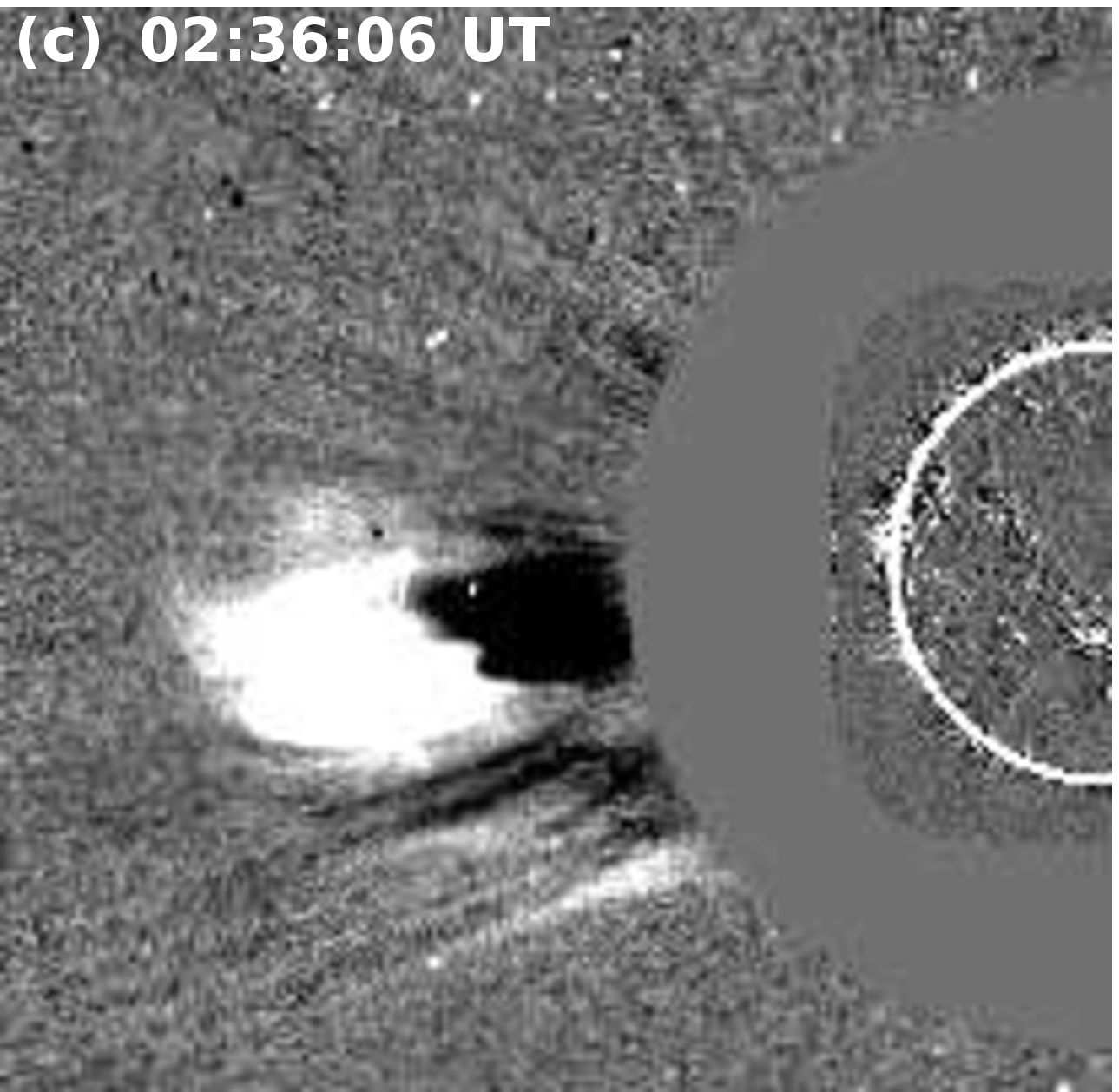}
}
\caption{(a) AIA 304 \AA~ image showing the C4.2 flare and associated jet. (b-c) CME associated with EUV jet was observed by SOHO/LASCO C2 coronagraph.}
\label{lasco}
\end{figure*}

\begin{figure*}
\centering{
\includegraphics[width=7cm]{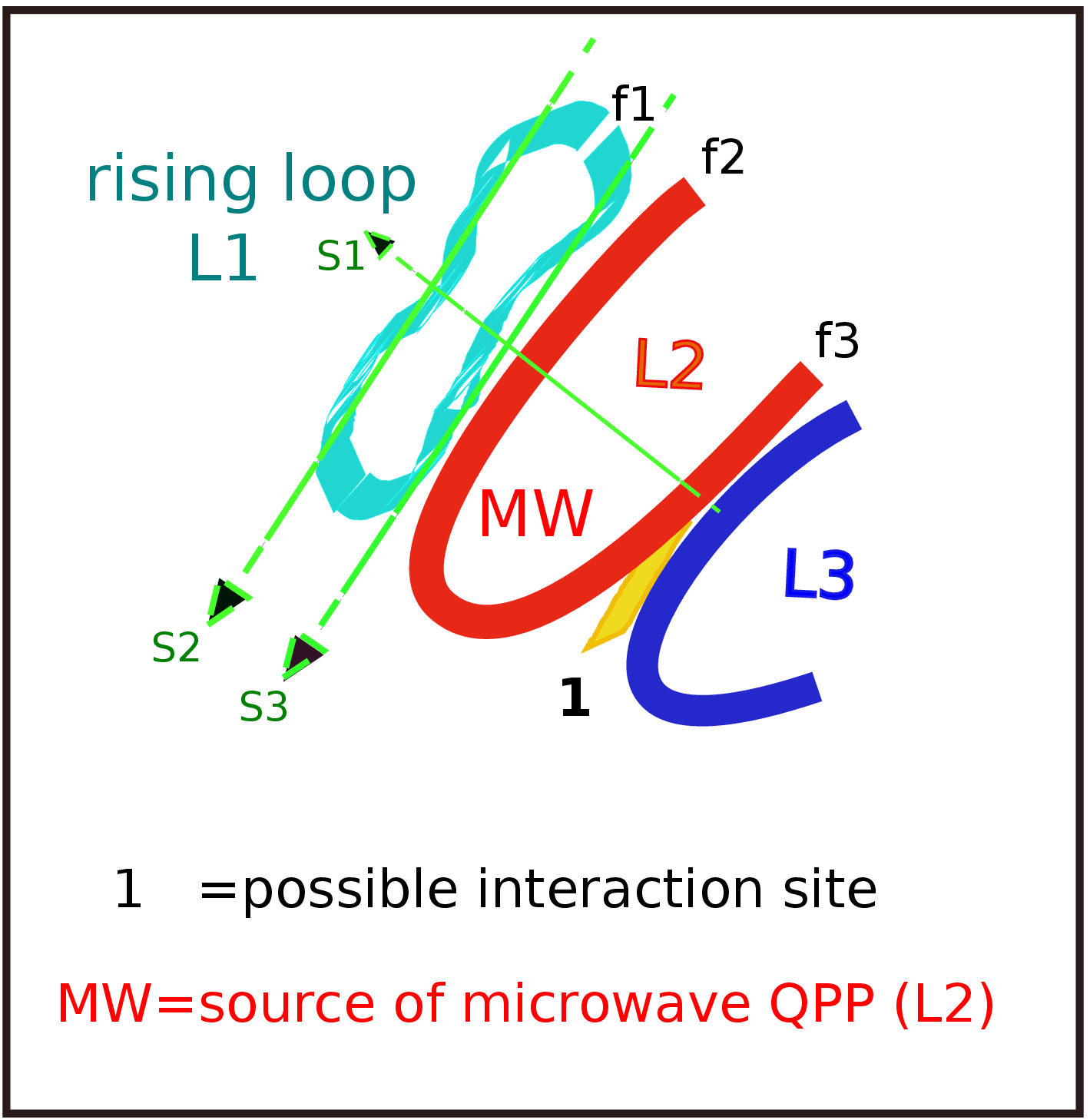}
\includegraphics[width=7cm]{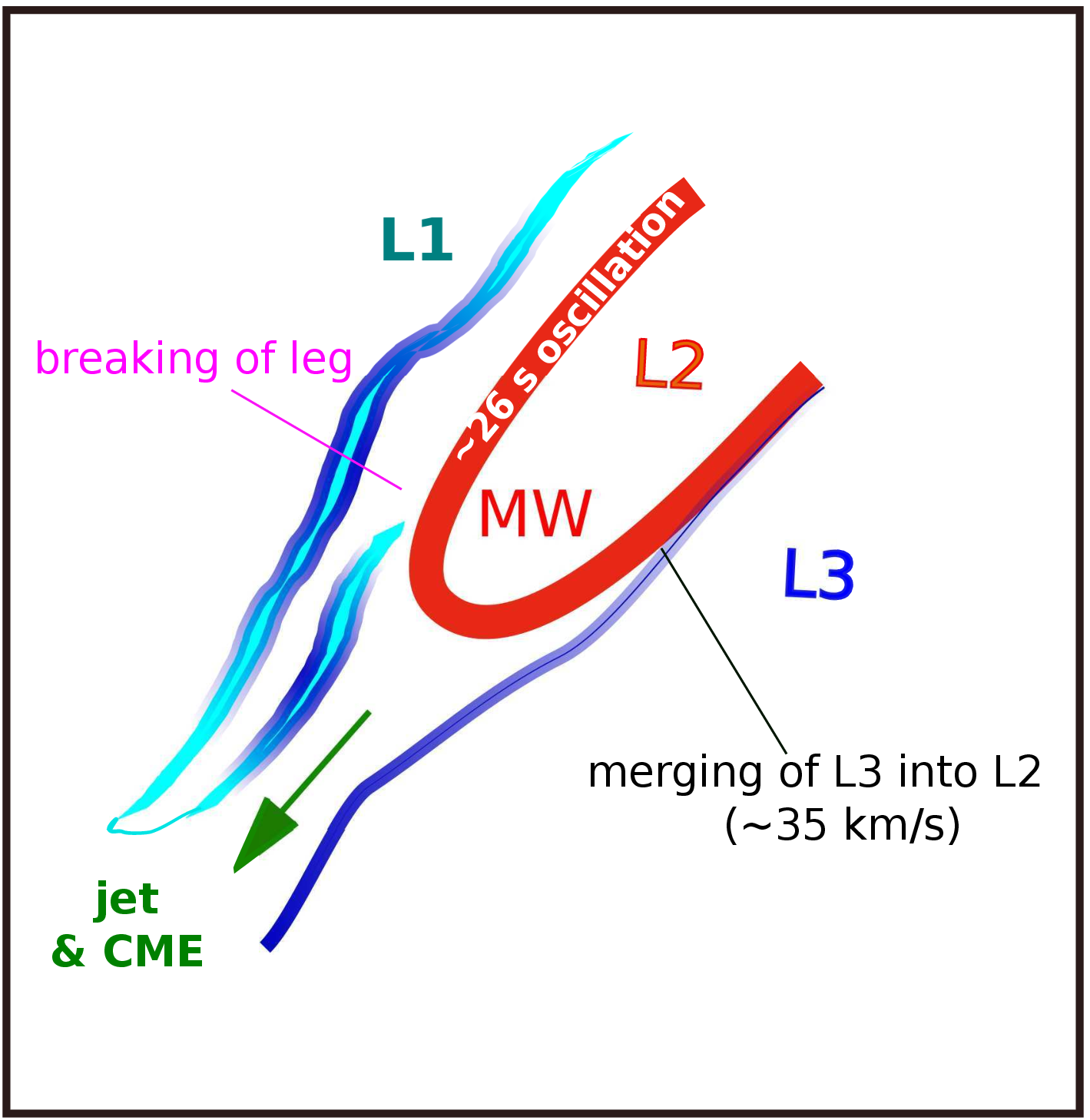}
}
\caption{Schematic cartoon showing the possible trigger of QPP in terms of loops' interaction. L1 is the rising loop. Loop L3 merges into L2 during the flare. Loop L2 is the main source of the microwave QPP. f1, f2, f3 are the footpoints (bright kernels) observed in the AIA 1600/1700 \AA~ channels. S1, S2, S3 indicate the slice cuts used to create the time-distance intensity plots using AIA 304 \AA~ images.}
\label{cart}
\end{figure*}

\section{OBSERVATIONS AND RESULTS}
The {\it Atmospheric Image Assembly} (AIA; \citealt{lemen2012}) onboard the
{\it Solar Dynamics Observatory} (SDO) records
full disk images of the Sun (field-of-view $\sim$1.3 R$_\odot$) with a
spatial resolution of 1.2$\arcsec$ (0.6$\arcsec$ pixel$^{-1}$) and the time cadence
of 12~s. For the present
study, we utilised the sequences of images taken at 304 \AA~ (\ion{He}{2}, T$\approx$0.05 MK), 171~\AA\ (\ion{Fe}{9}, $T\approx$0.7
MK), 94~\AA\ (\ion{Fe}{10},\ion{Fe}{18}, $T\approx$1 MK, $T\approx$6.3 MK), 131~\AA\ (\ion{Fe}{8}, \ion{Fe}{21}, \ion{Fe}{23}, i.e., 0.4, 10, 16 MK),  1600~\AA\ (\ion{C}{4} + continuum,
$T\approx$0.01 MK), and 1700~\AA\ (continuum, 5000 K). We also used the  Helioseismic and Magnetic Imager (HMI, \citealt{schou2012}) magnetograms to determine the magnetic topology of the AR. We utilised the Fermi Gamma-ray Burst Monitor (GBM, \citealt{meegan2009}), the Nobeyama Radioheliograph (NoRH, \citealt{nakajima1994}) and Radio Polarimeters (NoRP, \citealt{nakajima1985}) observations to investigate the HXR and microwave emissions in different energy bands. 

On 21 September 2015, AR NOAA 12420 (of the $\beta\gamma$ magnetic configuration, S08E80) was located close to the eastern limb. The QPP, reported here, was associated with an impulsive/compact C4.2 flare that started in this AR at  about 01:57~UT, reached its peak at about 02:04~UT, and ended at
02:08~UT. The QPP was observed during 02:01:45-02:02:45~UT, for about one minute.

\subsection{X-ray and radio observation of the QPP}
The left panel of Fig.~\ref{fermi_rstn} displays the Fermi GBM 4-s cadence X-ray flux profiles in different energy bands. The QPP was observed in the HXR channels, from 12--25 keV up to 100--300 keV energy. It suggests the periodic acceleration of non-thermal electrons up to 300 keV energy. The 100--300 keV channel reveals a decaying oscillatory pattern with three consecutive bursts of the energy release.

The right panel of Fig.~\ref{fermi_rstn}) shows the Radio Solar Telescope Network (RSTN) 1-s cadence flux density profiles observed at the Learmonth observatory in 245, 2695, 4995, 8800, and 15400~MHz frequencies. Interestingly, the decaying oscillation is observed in the microwave, i.e., 4995, 8800, and 15400~MHz. The 245 MHz shows a single burst suggesting the type III radio burst excited by an electron beam escaping upward from the reconnection region into the interplanetary medium.

The left panel of Fig.~\ref{nobeyama} displays the NoRP flux profiles obtained at 3.75, 9.4, and 17~GHz channels. We see clearly four consecutive decaying peaks of the microwave emission in 9.4 GHz, while the 17~GHz flux profile is consistent with the RSTN 15.4~GHz flux. The top right panel shows the NoRH flux profile 34~GHz signal with the 1-s cadence. The decaying oscillatory pattern is clearly observed in the 34~GHz channel. Although the QPP is clearly seen in the original light curves, we detrended the original signal by subtracting a 70-s smoothed signal shown by the dashed blue line. We adopted wavelet analysis \citep{torrence1998} of the detrended light curve to determine the oscillation period. The bottom panel of Fig.~\ref{nobeyama} displays the wavelet power spectrum of the detrended 34~GHz signal and the 99\% confidence level contours. The oscillation period is estimated as about 13-s.    

It is evident that the decaying QPP was observed only in the hard X-ray and microwave channels, therefore, the periodic precipitation/trapping of the energetic electrons in the high-density loop structures should be the main source of the observed QPP.

\subsection{Spatial location of the microwave sources}
To determine the spatial location of the 17 and 34~GHz sources, we overplotted 17 and 34~GHz emission intensity contours over the HMI magnetogram and the AIA 304, 1600, and 1700~\AA~ images during the peak time of each burst (Fig.~\ref{cont}). The HMI magnetogram image at 02:01:18~UT (Fig.~\ref{cont}a) shows the distribution of photospheric magnetic polarities in the active region and location of the 17 and 34 GHz sources at 02:01:55~UT near the edge of a big sunspot.

Fig.~\ref{cont}~(b-d) display the AIA 304 \AA~(chromosphere/transition region) images during the observed QPP, which are overlaid by 17 and 34~GHz microwave sources at 02:01:55~UT, 02:02:07~UT, and 02:02:28~UT, respectively. We see a radially upward rising thick loop, L1, and the microwave sources lie over the loop L2 that was heated during the rising motion of loop L1.

The AIA 1600 and 1700 images show the rising structure, L1, and formation of three bright kernels f1, f2, and f3, which are the footpoints of loops L1 and L2 (Fig.~\ref{cont}~(e-f)). Footpoint f1 belongs to the rising loop L1 whereas f2 and f3 are the footpoints of loop L2. The microwave sources are located above footpoints f2 and f3.

\subsection{The trigger of the QPP}
To investigate the trigger mechanism of the QPP, we analyse AIA 304, 171, and 131~\AA~ images. Fig.~\ref{aia304_171} displays selected AIA 304 and 171~\AA~ images during the observed QPP. Interestingly, we see brightening associated with an expanding flux tube (L1) in the AIA 304~\AA~ images at 02:01:43~UT. The plasma along the loop looks like a sausage-like structure drawn in panel (b). Taking a cut along L1, we observe a bright-dark-bright pattern along L1. A similar expanding flux tube was also detected during a flare by \citet{srivastava2013}. They observed bright knots over an entire flux tube existing for about 60 s, suggesting it was a morphological signature of the sausage instability. However, we see a clear separation between the legs of L1 during its expansion associated with bright-dark patches along the flux tube legs.
During the rise of L1, we see appearance of another loop, L2, in the southward direction. It may be interpreted as the interaction of expanding loop L1 with a preexisting structures in the active region, which makes L2 brighter during the rise of L1. One of the legs of loop L1 disconnected during the rising motion, leading to a surge-like ejection of the plasma. We are still observing structure L2 while loop L1 got destroyed and produced the surge-like ejection (Fig.~\ref{aia304_171}~(h-i)). The structures similar to loops L1 and L2 are observed in the AIA 171 \AA~ images during the flare (Fig.~\ref{aia304_171}~(j-l)). The 13-s QPP was observed during the rising motion of L1. The loops L1/L2 were clearly observed in the AIA 304 and 171~\AA~ channels simultaneously, which seems to indicate that these loops are filled in with a dense chromospheric/transition region plasma.

We selected slice cuts S1, S2, and S3 (Fig.~\ref{aia304_171}(e)) to create time-distance intensity plots using the AIA 304 \AA~ images. The purpose of the stack plots is to show the dynamics of L1 and L2 during the QPP. Fig. \ref{st_304} displays the stack plots along S1, S2, and S3 during 02:00-02:07~UT. Cut S1 is taken across the loop, whereas S2 and S3 are along the rising loop L1. The NoRH 34~GHz flux and Learmonth dynamic radio spectrum are included to show the temporal evolution of L1 during the observed QPP. We see the simultaneous evolution (i.e., rise) of L1 and observed QPP in the 34~GHz channel. We noticed blob-like structures in the loop L1 whose morphology seems to be a sausage-like plasma structure. Furthermore, the QPP was observed during the expansion of loop L1 (Fig.~\ref{st_304}(a)). The expansion of the loop seems to cause a brightening of nearby loop L2. The interaction of L1 with the preexisting structures in the active region produced a cool jet seen at 304 and 171~\AA, directed along the open structures above the flare site. Fig.~\ref{st_304}(a) reveals an intensity oscillation in the legs of L2. The average smoothed counts extracted from  between two horizontal lines (see panel (a)) indicates the presence of a 26-s oscillation in the leg of loop L2 during the observed QPP. However, after the QPP the oscillation period seems to become longer. Unfortunately, we cannot detect the intensity oscillation with the period shorter than 26~s, because of the limited temporal resolution (12 s) of the AIA. The estimated speed of the jet, determined from the linear fit, is 106--118 \kms (Fig.~\ref{st_304}(b-c)).    

Also, we observed a repetitive type III radio burst (Fig.~\ref{st_304}(d)) that was co-temporal with the microwave QPP. It suggests the periodic injection of non-thermal electrons along the open field lines, accelerated in the vicinity of the magnetic reconnection site in the flare. The simultaneous observation of the type III radio, HXR and microwave bursts suggest the injection/precipitation of the electrons bidirectionally. The downward moving non-thermal electrons generate the HXR and microwave QPP, whereas the upward moving electrons generate the type III radio bursts. In addition, we noticed a jet-like successful eruption of loop L1, generating a narrow CME in the SOHO/LASCO C2 coronagraph images. 
 
Interestingly, the oscillations were also detected in the time derivative of the soft X-ray flux obtained with GOES at 1--8 and 0.5--4~\AA. The derivative signal was smoothed by 6~s, and detrended by subtracting a 40-s smoothed curve. The right panel of Fig.~\ref{st_304} displays the wavelet power spectrum of the smoothed/detrended time derivative of the GOES 1-8 \AA~ flux signal taken with the 2-s cadence, using the three-point (quadratic) Lagrangian interpolation technique. The wavelet spectrum reveals the 26-s periodicity above the 99\% confidence level. This oscillatory component persists for longer duration in the SXR flux derivative in comparison to the HXR/microwave emission. 

To study the evolution of the structures in the hot channel ($\sim$7--10 MK), we analysed the AIA 131 and 94~\AA\ images. Fig.~\ref{aia131_94} displays the AIA 131 and 94~\AA\  images taken during the flare onset, impulsive and decay phases (02:00:08--02:05:49 UT). Interestingly, we observe two bright loops (L2 and L3) during the flare initiation (Fig.~\ref{aia131_94}(a)). Later on, we see the rise of loop L1 (indicated by the arrow) near one of the footpoints of loop L2. We see the rising motion of L1 during the flare impulsive phase. In addition, we also observe the merging of loop L3 into L1. The contours of NoRH 17 (green) and 34 (blue) GHz sources on the AIA 131 \AA~ image suggest that the microwave emission is generated in loop L2 (Fig.~\ref{aia131_94}(c)). After the merging of L3 with L2, we see a single loop (AIA 94 \AA) during the flare decay phase (Fig.~\ref{aia131_94}(f)).

To investigate the time evolution of loops L1 and L3 with L2, we chose slice cuts S4 and S5 in the AIA 131 and 94~\AA~ images (Fig.~\ref{aia131_94}(b,e)). Fig.~\ref{st_131} shows the time-distance intensity plot along cuts S4 and S5 during 02:00--02:05~UT. We see the merging of L3 with L2. The approaching speed of loop L3 toward L2 was about 35~\kms. We overplotted the NoRH 34~GHz radio flux to show the evolution of the loop interaction with the observed QPP. The timing of the QPP coincided exactly during the rise of L1. However, we also see the interaction of L3 with L2 at the same time as the QPP. However, the location of the microwave sources suggests that loop L2 is the source of the QPP observed in the microwave emission. Very likely, the non-thermal electrons trapped in loop L2 generated the microwave emission.

\subsection{DEM analysis of the loops}
To determine the temperature and emission measure in the interacting loops, we performed differential emission measure (DEM) analysis using an automatic code developed by \citet{asc2013}. We utilised six near simultaneous AIA images taken in the 171, 335, 211, 193, 131, 94~\AA~ channels at about 02:01:59~UT. The co-alignment of AIA images was done by using the limb fitting method, with the accuracy less than 1 pixel. The code fits a DEM solution in each pixel, which can be fitted by a single Gaussian function that has three free parameters: the peak emission measure (EM$_p$), the  peak temperature (T$_p$), and the temperature width sigma ($\sigma_T$). The images are selected just before the QPP occurrence at HXR and microwave, to avoid the unwelcome artefacts due to the image saturation during the flare. The peak temperature (T$_p$) and emission measure maps (EM$_p$) are shown in Fig.~\ref{dem}. Loops L2 and L3 are clearly observed in the EM maps, suggesting the high plasma density in the loops. Interestingly, we do not see loop L2 to appear complete in the T map, as only a part of the loop near the footpoint region is seen. On the other hand, loop L3 is well observed and reveals the high temperature of about 6~MK. Loop L2 was observed in both the cool (304, 171~\AA) and hot (131 and 94~\AA) channels suggesting the presence of a multi-thermal plasma. On the other hand, loop L2 was best observed only in the hot (131, 94~\AA) channels, and the interaction of the loops was observed in the hot channels too. 

We determine the mean values of the peak T$_p$, EM$_p$, $\sigma_T$ by selecting a region (box) near the top of loops L2 and L3. We estimate the total emission measure (TEM) by the integral of the Gaussian DEM distribution over the entire temperature range. The average values of peak T and EM in the boxes are found to be about 1.58~MK, $2.2\times10^{22}$~cm$^{-5}$\,K$^{-1}$, respectively, for loop L2; and about 5.4 MK, $4.58\times10^{22}$~cm$^{-5}$\,K$^{-1}$, respectively, for loop L3. The estimated TEM are about $1.92\times10^{28}$ cm$^{-5}$ for loop L2, and $3.90\times10^{29}$~cm$^{-5}$ for loop L3.     
Using the values of TEM, we calculated the densities at the loop top region, assuming that the column depth of the structure along the line of sight is nearly equal to its width. The apparent widths ($w$) of loops L2 and L3 are about 4.5$\arcsec$ and 4.2$\arcsec$, respectively. Therefore, the densities of the selected regions of loops L2 and L3 are estimated with the use of $({\mathrm{TEM}}/{w})^{1/2}$ to be about $7.6\times10^{9}$ cm$^{-3}$ and $3.5\times10^{10}$~cm$^{-3}$, respectively.

\subsection{An Associated Coronal Mass Ejection}
 The cool jet-like ejection from the chromosphere was seen to produce a narrow CME that was observed by the {\it Large Angle and Spectrometric Coronagraph}  on the {\it Solar and Heliospheric Observatory} (SOHO/LASCO). Fig.~\ref{lasco} displays an AIA 304~\AA~ image taken after the QPP, showing a jet and the associated CME, and the following CME. According to the SOHO/LASCO catalogue, the first appearance of the CME was at 02:12:04~UT in the LASCO C2 field of view.
 The linear speed and acceleration of the CME were about 769~\kms and -8.4~m\,s$^{-2}$, respectively. The observation of a metric (Learmonth) and interplanetary type III (Wind/WAVES) radio burst during the QPP also indicates the motion of the accelerated particles along the open field lines during the flare.  


\section{DISCUSSION AND CONCLUSION}
We analysed a short period, 13-s QPP observed in HXR and microwave emissions during the impulsive phase of a short-duration C4.2 flare. The possible mechanisms for the QPP are discussed below.

We observed a radially rising small loop, L1, with a speed of about 100~\kms~ at the time of the flare initiation. During the loop moving from the chromosphere to corona, we observed breaking of one of its legs during its rising motion. Loop L1 may interact with the ambient preexisting field located south to L1. The interaction could be in a similar way, e.g., interaction of a small filament with preexisting open structures in the active region \citep{sterling2015, kumar2016}, which produces a cool jet. The breaking of the leg of L1 and simultaneous appearance/heating of L2 could be a result of the interaction of L1 with the preexisting structures in the active region. 
The AIA time-distance intensity plots reveal that the QPP is occurring at the time of the expansion and rising motion of L1. The SDO/AIA and NoRH 17 and 34~GHz observations suggest that loop L2 is the main source of the observed QPP in the microwave emission. Thus, the question is what triggers the QPP in loop L2. Furthermore, we also see the appearance of loops L2 and L3 in the hot AIA channels (131/94 \AA) before the rising motion of L1 and triggering of the microwave/HXR QPP. Therefore, the coalescence of L3 into L2 seems to be the most likely cause of the observed microwave QPP. 

The converging speed of loops L2 and L3 is about 35~\kms, which is consistent with the previous report ($\sim$30 \kms) of loop-loop interaction observed by \citet{kumar2010c}. In that study, the loop length was estimated as about 100~Mm, and the period of QPP was measured in the HXR and microwave channels as $\sim$40~s. Therefore, a fundamental standing wave, e.g. the kink or sausage mode of the loop, would have the phase speed of about 5000~\kms. However, it was not possible to detect a kink or sausage oscillation of the loop during the QPP directly, because of the lack of the necessary resolution. In our case, QPP is associated with a much smaller loop, L2. The distance between its footpoints was $\sim$18$\arcsec$ (AIA 94~\AA, Fig.~\ref{aia131_94}(e)) Assuming the semi-circular shape of the loop, we estimate the loop length as $\sim$20 Mm. For the 13-s period it gives us the phase speed ($2L/P$) of a fundamental mode as $\sim$3000 \kms. This estimation is consistent with the typical values of the fast magnetoacoustic speed in coronal active regions.   
Thus, the short period QPPs could be excited by kink or sausage fast magnetoacoustic oscillations. However, we do not observe any transverse oscillation of the loops at the flare site, and therefore the interpretation of the observed QPP in terms of MHD oscillations does not have any further observational support in the discussed case. But, the short period of the possible oscillation would mean the small transverse displacement of the loop, which could be smaller than the resolution of the instrument. Moreover, the detected period is about the AIA cadence time, which also makes the possible oscillation undetectable. In addition, According to the numerical simulation of a global sausage mode \citep{pascoe2007}, the density contrast of 5--10 requires a 20 Mm loop length to support a 13~s trapped sausage oscillation. The length of loop L2 is 20 Mm, and hence it would be able to generate the sausage oscillation with the observed period.

We also detected an intensity oscillation with the period of about 26~s in the legs of the flare loop (L2) observed in the AIA 304 \AA~ channel. The oscillation continued even after the HXR and microwave QPP. This oscillation could also be interpreted as a magnetohydrodynamic mode of loop L2. 
In addition, AIA 304 \AA~ and GOES SXR derivative signals reveal increase in the oscillation period from 26~s to 35-40~s during the flare maximum (02:03--02:05:30~UT). If the oscillation is caused by the sausage mode, the increase in the period may be explained by the increase in the density of the flare loop caused by the chromospheric evaporation after the precipitation of nonthermal electrons. For the plasma beta $\beta<$1 and  a constant magnetic field, the period of the sausage mode depends on the  density of the loop (P$\sim$$\sqrt{n}$, \citealt{stepanov2012}). This scenario favours global sausage-mode oscillation of the flare loop as a driver of the 26-s QPP. In this case the 13-s QPP in HXR/microwave could be interpreted as a second longitudinal harmonic of the sausage mode, with the wavelength equal to the loop length. Another option is that the 26-s oscillation is a subharmonic of the 13-s oscillation, pumped by the parametric resonance \citep[e.g.,][]{2008CosRe..46..301Z}.

On the other hand, the interaction of the loops observed in this event may lead to the oscillatory reconnection and acceleration of nonthermal electrons from the loop coalescence site to the footpoint region, producing therefore the QPP detected in the microwave and HXR emissions. The loop coalescence model \citep{tajima1987,sakai1987} suggests the presence of (i) quasi-periodic oscillations of fields and the number of accelerated charged particles (ii) double peak structure of the oscillation profile.
In our case, we indeed observed a double peak structure in the first HXR burst in the 12--25~keV, 25--50~keV, 50--100~keV energy channels, and in the first peak of the 9.4~GHz radio flux. Moreover, in loop--loop interaction events, the loops' morphological changes are expected, that result usually in confined flares. Indeed, in this event we observed a complete disruption of loop L3, and a cool jet-like eruption after merging of L3 into L2. The observation of type III radio bursts (25--180~MHz) during the QPP indicates the injection of energetic electrons upward from the reconnection site along the open field lines. Therefore, the simultaneous observation of QPP in hard X-ray, microwave and type III radio burst suggests the quasi-periodic injection of the non-thermal electrons bidirectionally (e.g., \citealt{kumar2016}).  Thus, our observation suggests that the oscillatory regime of the loop--loop interaction is the likely driver of the detected QPP. 

Recently, \citet{mes2016} reported a sub-second pulsation (0.07--1.49 s period) in broadband microwave (4--7 GHz) emission for 70~s during a C-class flare. The pulsation was associated with an expanding hot loop observed in the AIA 131, 94, and 193~\AA~ channels. They did not observe any cool plasma ejection or jet in the AIA 304 or 171~\AA~ channels. However, in our case, we see a clear coalescence of loop L3 into L2 observed in the AIA hot channels (131 and 94 \AA), which is the most likely cause of the observed QPPs.

The possible scenario of the QPP drivers in terms of the oscillatory coalescence of loops L2, and L3, is shown by a schematic cartoon (Fig.~\ref{cart}). Loop L3 is destroyed by the interaction with L2, and produces a narrow jet-like eruption. The microwave emission coming from L2 is possibly produced by the trapped electrons accelerated during the reconnection. The HXR emission is expected to be originated from the footpoints (f2/f3) of loop L2. 

In conclusion, we report a 13-s QPP in the non-thermal emission in the HXR and microwave channels, and a 26-s QPP in the thermal emission, in the EUV and soft X-ray flux derivative. In this event, we could observe a clear interaction of loops (merging of L3 into L2), which most likely generated a 13-s QPP by the periodic acceleration of nonthermal electrons, produced by the oscillatory coalescence of flaring loops. The sausage oscillation of flare loop L2 cannot be rule out completely and may also be a possible candidate to generate the QPPs. The 26-s periodicity of the thermal emission may either result from the parametric resonant pumping by the 13-s oscillation, or be a sausage oscillation excited independently of the 13-s oscillation. In the latter case the 13-s and 26-s periodicities could match each other by accident.
Future analysis of high-resolution multiwavelength data sets will shed more light on the generation mechanisms of QPPs.   
\acknowledgments
SDO is a mission for NASA Living With a Star (LWS) program. This work was supported by the \lq\lq Development of the Korea Space Weather Center\rq\rq\ of KASI and the KASI basic research funds. The SDO data was (partly) provided by the Korean Data Center (KDC) for SDO in cooperation with NASA and SDO/HMI team. The Nobeyama RadioHeliograph and Polarimeters are operated by the International Consortium for the Continued Operation of Nobeyama.  KSC acknowledges support by a grant from the US Air Force Research Laboratory, under agreement number FA 2386-14-1-4078 and by the \lq\lq\ Planetary system research for space exploration\rq\rq\ from KASI.
VMN acknowledges the support from the European Research Council under the \textit{SeismoSun} Research Project No. 321141, and the BK21 plus program through the National Research Foundation funded by the Ministry of Education of Korea. 
Wavelet software was provided by C. Torrence and G. Compo, and is available at \href{http://paos.colorado.edu/research/wavelets/}{http://paos.colorado.edu/research/wavelets/}.
 This work was (partly) carried out on the Solar Data Analysis System operated by the Astronomy Data Center in cooperation with the Solar Observatory of the National Astronomical Observatory of Japan.


\bibliographystyle{apj}
\bibliography{reference}
\clearpage

\end{document}